\documentclass[preprint,onecolumn,amsmath,amssymb,floatfix]{revtex4-2}

\usepackage{color}
\usepackage{mathrsfs}
\usepackage{graphicx}
\usepackage{dcolumn}
\usepackage{bm}
\usepackage{hyperref}
\usepackage{amsmath,amsfonts,amssymb,ulem}
\usepackage{epstopdf}
\usepackage{xcolor}
\usepackage{multirow}
\usepackage{newtxtext,newtxmath}

\usepackage{graphicx}

\begin{document}
\title{
Enhancing 10 TeV $\gamma\gamma$-collider luminosity through scattering-laser wavelength selection in the presence of prolific electron-positron pair production
}
\author{S. S. Bulanov}\email[]{sbulanov@lbl.gov}
\affiliation{Lawrence Berkeley National Laboratory, Berkeley, California 94720, USA}
\author{T. Barklow} 
\affiliation{SLAC National Accelerator Laboratory, Menlo Park, California, USA}
\author{C. Benedetti}
\affiliation{Lawrence Berkeley National Laboratory, Berkeley, California 94720, USA}
\author{A. Formenti}
\affiliation{Lawrence Berkeley National Laboratory, Berkeley, California 94720, USA}
\author{S. Gessner}
\affiliation{SLAC National Accelerator Laboratory, Menlo Park, California, USA}
\author{R. Lehe}
\affiliation{Lawrence Berkeley National Laboratory, Berkeley, California 94720, USA}
\author{A. Rastogi}
\affiliation{Lawrence Berkeley National Laboratory, Berkeley, California 94720, USA}
\author{S. Pagan Griso}
\affiliation{Lawrence Berkeley National Laboratory, Berkeley, California 94720, USA}
\author{C. B. Schroeder}
\affiliation{Lawrence Berkeley National Laboratory, Berkeley, California 94720, USA}
\author{A. Schwartzman}
\affiliation{SLAC National Accelerator Laboratory, Menlo Park, California, USA}
\author{J. Osterhoff}
\affiliation{Lawrence Berkeley National Laboratory, Berkeley, California 94720, USA}

\begin{abstract} 
A $\gamma\gamma$-collider capable of reaching the 10 TeV parton-center-of-momentum (pCM) frontier of particle physics may enable the study of phenomena beyond the Standard Model. Based on compact linear wakefield accelerator technology, such a collider could be realized by Compton scattering multi-TeV lepton beams off moderate-intensity laser pulses close to the collider interaction point, producing the required $\gamma$-photons. It is shown that, for a wide range of scattering-laser wavelengths, $\gamma\gamma$-collisions at the interaction point can meet the luminosity requirements for novel particle physics studies, even in the presence of the prolific electron-positron pair production that accompanies the interaction of the scattering laser with the multi-TeV lepton beam. Notably, this pair production imposes a natural limit on the maximum achievable photon luminosity. Accounting for this limit and for the angular divergence of the Compton photons yields an enhanced $\gamma\gamma$-collider luminosity for 250 nm and 1.25 nm scattering lasers. Moreover, the secondary pairs can themselves be exploited in physics studies, since their luminosity is high enough to produce heavy particles at rates needed for discoveries well beyond the reach of existing colliders.
\end{abstract}

\maketitle

\section{Introduction}

The scientific community has identified \cite{P5:2023report} the study of the 10 TeV pCM energy frontier as a priority for exploring the fundamental building blocks of matter beyond the predictions of the Standard Model of particle physics \cite{weinberg.prl.1967}. Research at 10 TeV pCM will require novel collider and accelerator concepts for hadron or lepton beams, as no technology available today is ready to realize a collider at this energy scale. While hadron and heavy-lepton accelerators can be circular \cite{FCC_hh.arxiv.2022,muoncollider.arxiv.2025}, electron-positron ($e^+e^-$) accelerators must be linear \cite{CLIC-CDR.cern.2012,ILC.arxiv.2013} because of the limitations posed by synchrotron radiation. Since radio-frequency (rf) accelerator technologies reach the limit of their economic viability at multi-TeV energies for linear machines, a number of new particle-beam acceleration concepts are being developed, including wakefield-based schemes \cite{schroeder.jinst.2023}. Wakefield acceleration \cite{esarey.rmp.2009, jing.rast.2016, lindstrom.arxiv.2025} is under active development and promises to miniaturize the accelerator by supporting GV/m accelerating fields, outperforming rf systems by more than an order of magnitude. Recent progress has been notable and includes the demonstration of 10 GeV electron beams from a laser-driven accelerator \cite{picksley.prl.2024} using a novel 30 cm plasma source \cite{miao.prx.2022}, emittance preservation in beam-driven plasma accelerators \cite{lindstrom.natcomm.2024}, and the demonstration of high transformer ratios in dielectric wakefield accelerators \cite{gao.prl.2018}. Three different configurations are being investigated for a wakefield-based energy-frontier accelerator, enabling $e^+e^-$, electron-electron ($e^-e^-$), and $\gamma\gamma$ collisions \cite{gessner.arxiv.2025}. An important requirement for particle colliders at the multi-TeV scale is extreme beam luminosity, which is needed to achieve statistically relevant production rates of the particles under study. This requirement translates into very intense colliding beams that generate strong electromagnetic fields; at the interaction point, these fields cause energy loss into photon radiation (the so-called beamstrahlung) and $e^+e^-$ pair generation \cite{schroeder.prab.2010,esberg.prab.2014,yakimenko.prl.2019}. As a result, the luminosity spectra become broad and the total luminosity at the highest collision energy can decrease. All configurations for a 10 TeV energy-frontier wakefield collider must therefore take strong beamstrahlung into account when estimating their ultimate physics reach and power requirements \cite{gessner.arxiv.2025}. 

Whereas the $e^+e^-$ and $e^-e^-$ configurations collide the primary accelerated beams, $\gamma\gamma$-colliders require an additional step to generate the photons. For this reason, ever since discussion of the successors to LEP II \cite{lep2.pr.2013} began, $\gamma\gamma$-colliders have been regarded as an important but complementary component of $e^+e^-$ collider proposals, including Higgs factory designs \cite{ginzburg.nim.1983, telnov.nima.1990,telnov.nima.1995,telnov.nima.2000,telnov.npb.2000, asner.epjc.2001, telnov.nima.2001a, saldin.nima.2001, sapphire.arxiv.2012, NLC_ZDR.2018, barklow.jinst.2023b}. Studies suggest that at 10 TeV the $e^+e^-$ option could offer the strongest physics case \cite{chigusa.arxiv.2025,cipressi.arxiv.2026,fraser.arxiv.2026}. This option, however, faces two challenges that pose a risk and could significantly delay the readiness window of such a collider: the absence of positron sources intense enough to provide the luminosities required at 10 TeV, and the lack of test-facility access and of high-efficiency concepts for positron acceleration in plasma wakefields \cite{musumeci.arxiv.2022,cao.prab.2024}. The $e^-e^-$ option, while technologically more mature, is considerably less attractive from a particle physics perspective, offering a reduced physics reach compared with both the $e^+e^-$ and $\gamma\gamma$ options under comparable conditions.

This opens a window of opportunity for a standalone 10 TeV $\gamma\gamma$-collider powered by two electron accelerators. Such a collider could enable a comprehensive physics program and support discoveries beyond Standard Model phenomena. In particular, any extension of the Standard Model that involves particles interacting under the electroweak force must contain at least one electrically charged particle. Such particles can therefore always be produced in the analogue of the Breit-Wheeler process ($\gamma\gamma \to X^+X^-$), regardless of the details of the model in question. Moreover, the inclusive $\gamma\gamma \to WW$ cross section is as large as $\sim 90$~pb (compared with $\sim 0.06$~pb for $e^+e^- \to WW$) \cite{tesla.ijmpa.2004}, which could present an opportunity for precision $W$ physics. The large $WW$ cross section suggests similarly enhanced cross sections for $\gamma\gamma\to WWX$, where $X$ denotes a set of beyond-the-Standard-Model particles that couple primarily to the $W$ and $Z$ bosons. Examples include additional Higgs bosons and weakly interacting massive particles (WIMPs), which are good dark matter candidates. The inclusive $\gamma\gamma \to WWh$ and $\gamma\gamma \to WWhh$ cross sections exhibit the same behavior, being $\sim 20$~fb and $\sim 0.5$~fb respectively, i.e., 400 and 250 times larger than those for $e^+e^- \to WWh(h)$. This might enable studies of the Higgs self-coupling and of anomalous gauge couplings. Both $e^-e^-$ and $\gamma\gamma$ collisions necessarily lead to the appearance of $e^+e^-$ pairs in the colliding beams. As we show below, the $e^+e^-$ luminosity of these secondary collisions in the $\gamma\gamma$ operational scenario is enhanced to the point that the production of heavy particles, $e^+e^-\to XX$, becomes viable and is more significant than in an $e^-e^-$ collider. Finally, a $\gamma\gamma$-collider would have unique sensitivity to heavy particles produced in $\gamma\gamma$ fusion \cite{RebelloTeles:2023uig}. Taken together, these considerations show that detailed investigations comparing the physics discovery potential of 10 TeV $\gamma\gamma$-colliders with that of other facilities are required. Recent studies \cite{chigusa.arxiv.2025,cipressi.arxiv.2026,fraser.arxiv.2026} have begun to evaluate the discovery capabilities of new collider configurations, including $e^+e^-$, $e^-e^-$, and $\gamma\gamma$ options at 10 TeV pCM, and have found that broad luminosity spectra do not undermine the electroweak physics case. Indeed, beam-beam effects, i.e., beamstrahlung and $e^+e^-$ pair production, which are responsible for beam particle energy loss and for broad luminosity spectra, can be beneficial for new particle discovery and for searches for resonant production channels.

\begin{figure*}[!ht]
\centering
\includegraphics[width=15.0cm]{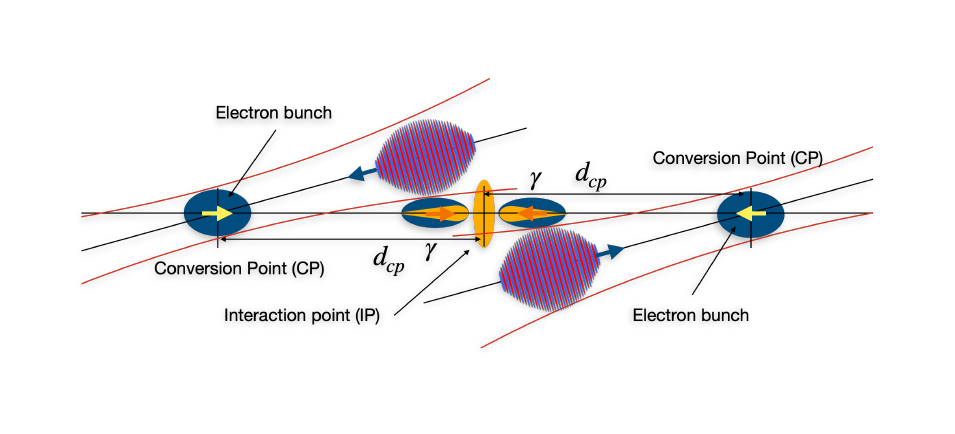}
\caption{Generic scheme of a $\gamma\gamma$-collider: two electron beams scatter off two laser pulses at the conversion points (CP), producing high-energy $\gamma$-beams that collide at the interaction point (IP).}
\label{fig:scheme}
\end{figure*}

In this paper, we study the optimization of the $\gamma$ production process and of the $\gamma\gamma$ interaction for a future 10 TeV pCM energy-frontier collider enabled by two wakefield-based electron accelerators \cite{schroeder.jinst.2023, gessner.arxiv.2025}. The scheme operates in two stages: first, the two multi-TeV electron beams each scatter off a separate laser, producing high-energy $\gamma$-photons through Compton scattering; second, these $\gamma$-beams, together with the respective electron beams, meet at the interaction point (IP) of the collider. It is well known that the efficiency of $\gamma$-beam generation in this way depends strongly on the laser parameters, including the intensity $I_0$, wavelength $\lambda_0$, and duration $\tau_0$, as well as on the distance $d_{CP}$ between the conversion point (CP), where the electrons and the laser collide, and the IP. A generic scheme of such a collider is shown in Fig.~\ref{fig:scheme}.

It is typically assumed that a $\gamma\gamma$-collider should use laser pulses with a wavelength chosen so as to prevent the generated high-energy photons from converting into $e^+e^-$ pairs via the Breit-Wheeler process during propagation through the laser pulse \cite{telnov.npb.2000,telnov.nima.2001a}. The governing parameter is $x=4E_0\hbar \omega_0/(m_e^2c^4)$, where $E_0$ is the electron energy, $\omega_0$ is the laser frequency, $m_e$ is the electron mass, $\hbar$ is the reduced Planck constant, and $c$ is the speed of light. The maximum photon energy emitted in Compton scattering is $\hbar\omega^\prime_{m}=[x/(1+x)]E_0$. Suppressing pair production requires $x<4.8$, which places a lower limit on the scattering laser wavelength, $\lambda[{\rm \mu m}]>3.96\,E_0[{\rm TeV}]$. For $\gamma\gamma$-collider Higgs factory designs \cite{barklow.jinst.2023b}, this limit leaves the laser wavelength comfortably in the optical range: for $E_0=250$~GeV and $\hbar\omega_0=1.17$~eV ($\lambda=1.06$~$\mu$m), one obtains $x=4.5$ and $\hbar\omega_m^\prime=0.82E_0=205$~GeV. For a 10 TeV collider, however, restricting $x<4.8$ requires two 6 TeV electron beams and a laser of $\sim 25$~$\mu$m wavelength, which poses a significant technological challenge.

It has recently been pointed out \cite{barklow.jinst.2023a,barklow.jinst.2023b} that using high-frequency lasers in the X-ray range may be beneficial for the $\gamma\gamma$-collider design despite the prolific pair production that accompanies them. For $x\gg 1$ (e.g., for a Higgs factory, a 62.8~GeV electron beam and a 1~keV laser pulse give $x\approx 1000$), the high-energy part of the Compton photon spectrum was found to be significantly enhanced, which in turn enhances the $\gamma\gamma$ luminosity near the maximum center-of-mass energy. For such X-ray lasers, however, the required pulse energy ($\sim 700$~mJ \cite{barklow.jinst.2023b}) lies well beyond the state of the art ($\sim 4$~mJ \cite{altarelli.tdr.2006}) and is expected to be higher still in the 10 TeV case. It is therefore prudent to explore scattering lasers of intermediate wavelength. These will still operate in the presence of significant pair production and retain a $\gamma\gamma$ luminosity enhancement, while the technological challenges become easier to address as the wavelength approaches and enters the optical range. To determine whether such a $\gamma\gamma$-collider concept at 10 TeV pCM can generate $\gamma\gamma$ luminosities that meet the requirements of high-energy physics studies, we explore the parameter space of laser wavelength, duration, and intensity. 

The paper is organized as follows. In Sec.~II we present an analytical study of high-energy photon generation in the interaction of a 5 TeV electron beam with a scattering laser pulse in the presence of $e^+e^-$ pair production. In Sec.~III we discuss computer simulations of the conversion point and the interaction point of the $\gamma\gamma$-collider, and we study the dependence of the various collision luminosities on the scattering laser parameters. We conclude in Sec.~IV. Appendix I discusses the effect of electron beam polarization on the Compton and Breit-Wheeler processes and on the resulting $\gamma\gamma$ luminosity spectrum, and Appendix II examines the dependence of the $\gamma\gamma$ luminosity on the electron bunch length. Appendix III illustrates the effect of a higher electron bunch charge, i.e., of an increased geometric luminosity, on the $\gamma\gamma$ luminosity spectrum, and Appendix IV gives a short overview of the {\sc cain} code and summarizes the electron beam and laser parameters used in the simulations.

\section{High energy photon generation in the presence of $e^+e^-$ pair production}

\begin{figure}[!ht]
\centering
\includegraphics[width=8.6cm]{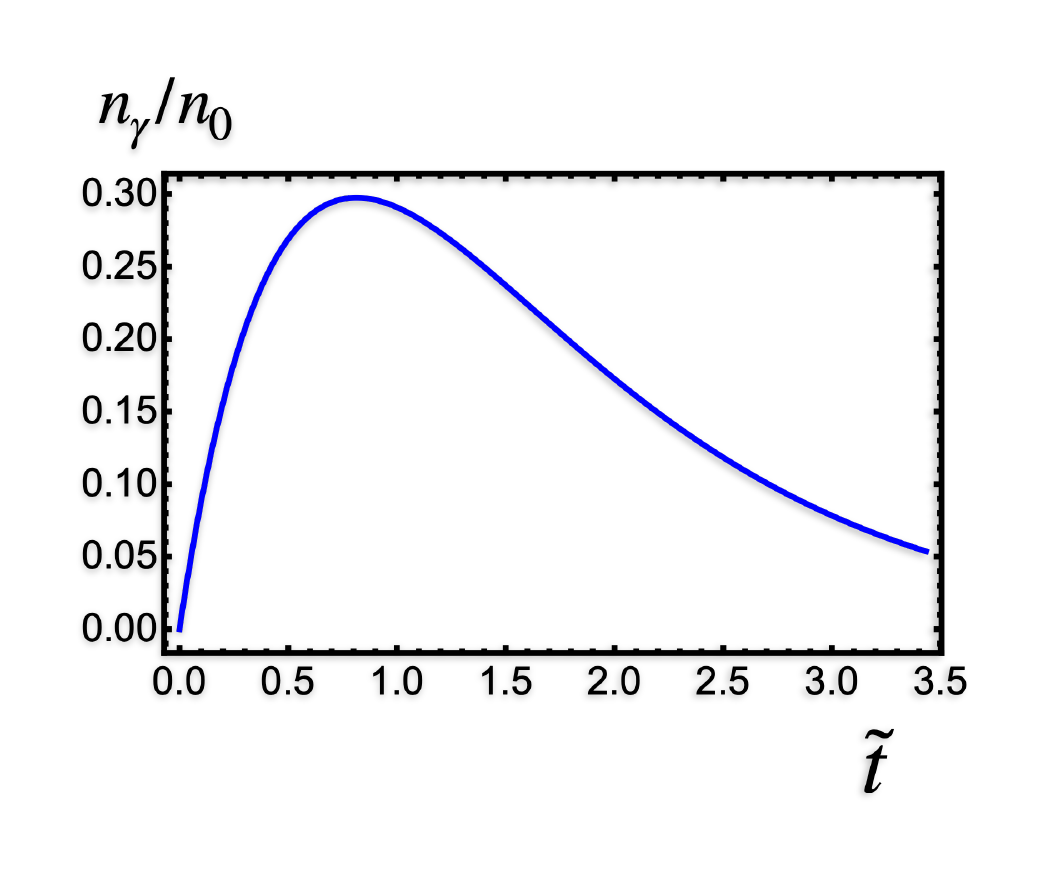}
\caption{Number of high-energy photons, normalized to the number of electrons in the beam, as a function of the laser pulse duration. The pulse duration is given in units of the electron mean lifetime with respect to the Compton process. Here $x=191$.}
\label{fig:number of photons}
\end{figure}

We first address the question of why the interaction of an electron beam with a scattering laser pulse can produce a sizable number of high-energy photons in the presence of prolific $e^+e^-$ pair production. Consider an idealized one-dimensional (1D) problem in which an electron beam interacts with a laser. If the laser intensity is sufficiently low, one can neglect the effects of the Lorentz force and retain only the linear Compton and Breit-Wheeler processes. We further assume that each electron emits at most one high-energy photon, where a ``high-energy'' photon is defined as one whose energy exceeds half of the initial electron energy. The evolution of the number of photons is then described by
\begin{equation}\label{eq:ne}
    n_e^\prime(t)=-n_e(t) W_C,
\end{equation}
\begin{equation}\label{eq:ngamma}
    n_\gamma^\prime(t)=n_e(t) W_C - n_\gamma(t) W_{BW},
\end{equation}
where the prime denotes differentiation with respect to time, $n_e$ is the number of electrons that have not yet emitted a high-energy photon, and $n_\gamma$ is the number of high-energy photons; the emission of photons by the electrons and positrons created in the Breit-Wheeler process is neglected. The rates of the Compton ($W_C$) and Breit-Wheeler ($W_{BW}$) processes are assumed to be constant. Equation~(\ref{eq:ne}) thus governs the depletion of the initial electrons, while Eq.~(\ref{eq:ngamma}) governs the evolution of the number of high-energy photons, which grows through Compton scattering and decays through the Breit-Wheeler process. The solution of Eqs.~(\ref{eq:ne}) and (\ref{eq:ngamma}) is
\begin{equation}
    n_e=n_0 e^{-\tilde{t}},~~n_\gamma=n_0(\kappa-1)^{-1}\left(e^{-\tilde{t}}-e^{-\kappa\tilde{t}}\right), \label{eq:nphoton}
\end{equation}
where $\kappa=W_{BW}/W_C$, $\tilde{t}=W_C t$, and $n_e(0)=n_0$ is the initial number of electrons. The number of photons reaches its maximum at
\begin{equation}
\tilde{t}_{max}=\ln(\kappa)/(\kappa-1) 
\label{eq:tmax}
\end{equation}
and this maximum is
\begin{equation}
   n_\gamma^{max}=n_0\kappa^{\frac{\kappa}{1-\kappa}}.
   \label{eq:nmax}
\end{equation}
Figure~\ref{fig:number of photons} shows that the number of photons grows at small $\tilde{t}$, since enough electrons are initially available to overcompensate the photon loss to pair production. After some time the two processes balance exactly and the number of photons reaches its maximum. For the typical laser and electron beam parameters used below, $\hbar\omega_0=2.5$~eV, $a_0=0.3$, and $E_0=5$~TeV, the parameter $x$ is $x=191$, which yields $n_\gamma^{max}=0.3n_0$ and $\tilde{t}_{max}=0.8$. The largest number of high-energy photons that can be generated in the presence of pair production is therefore approximately one per three initial beam electrons, obtained with a laser pulse of duration corresponding to $\tilde{t}_{max}$. Since the luminosity scales as $n_\gamma^2$, it is potentially reduced by a factor of 10 relative to the geometric luminosity of colliding electron beams,
\begin{equation}
    L_{geo}^{ee}\approx\frac{f n_0^2}{4\pi\sigma_r^2},
\end{equation}
where $f$ is the collision frequency and $\sigma_r$ is the bunch radius at the interaction point. Because $\kappa\rightarrow 2$ as $x\rightarrow\infty$, the maximum number of high-energy photons tends to $0.25n_0$ in this limit, corresponding to a luminosity reduction by a factor of 16. This 1D model of high-energy photon production in the collision of an electron beam with a laser pulse, retaining the two competing processes of photon emission and $e^+e^-$ pair production, thus predicts that the ratio of the $e^-e^-$ geometric luminosity to the $\gamma\gamma$ geometric luminosity (with the $\gamma$-beam divergence from Compton scattering not taken into account) varies from 1 to 16 as $x$ varies from 4.8 to $\infty$. 

This simple 1D analysis shows that a significant number of high-energy photons per initial electron can be generated in the presence of pair production, i.e., for $x>4.8$. As noted above, the inclusive cross section for $\gamma\gamma\rightarrow WW$ is three orders of magnitude larger than that for $e^+e^-\rightarrow WW$, which may more than compensate for the decrease in relative luminosity.

A second effect also reduces the $\gamma\gamma$ luminosity. The energy of an emitted photon and its emission angle are connected through four-momentum conservation, so photons with energy above a given value, $\hbar\omega^\prime>b\hbar\omega_m^\prime$ with $0<b<1$, are emitted within the angle $\theta_b\approx\sqrt{x(b^{-1}-1)}/\gamma$. This angle, together with $d_{CP}$, determines the radius of the high-energy photon beam at the interaction point, so the resulting luminosity degradation depends on both quantities. One can define a characteristic value of $d_{CP}$ at which the radius of the photon beam with $\hbar\omega^\prime>b\hbar\omega_m^\prime$ is doubled at the interaction point,
\begin{equation}
    d_{CP}=\frac{\sqrt{\epsilon\beta}}{\theta_b}, \label{dcp}
\end{equation}
where $\epsilon$ and $\beta$ are the geometric emittance and the beta function of the electron beam. Unless stated otherwise, we use $\beta=0.6$~mm and $\epsilon=50/\gamma$~nm, chosen so that the Oide limit \cite{oide.prl.1988,blanco.prab.2016} does not dominate the beam focus size. For photons with energy $\hbar\omega^\prime=0.8E_0$, the characteristic $d_{CP}$ ranges from 0.12~mm for a 1~keV laser to 3.7~mm for a 1.2~eV laser. Maximizing the luminosity of a $\gamma\gamma$-collider therefore requires generating the maximum number of photons, Eq.~(\ref{eq:nmax}), over a laser pulse length corresponding to $t_{max}$, Eq.~(\ref{eq:tmax}), with the CP-IP distance given by Eq.~(\ref{dcp}). Assuming $d_{CP}=c\tau/2$, so that the electron beam interacts with the entire laser pulse before reaching the IP, the solution of $ct_{max}=2\sqrt{\epsilon\beta}/\theta_b$ gives the laser photon energy at which both conditions are met simultaneously. For $b=0.95$, this yields $\hbar\omega_0\approx 10$~eV and $c\tau\approx 5$~mm, placing a possible optimized operating point of a $\gamma\gamma$-collider at a laser wavelength of $\lambda\sim 100$~nm and a pulse length of 5~mm. This 1D analysis does not, however, account for the effects of electron beam polarization, for the enhancement of high-energy photon production with increasing $x$, or for the contributions of beamstrahlung and pair production at the interaction point. Computer modeling of the interaction is required to include these effects.   

\section{Modeling a 10 TeV $\gamma\gamma$-collider}

In what follows we report on simulations of the electron beam interaction with the scattering laser at the conversion point and of the subsequent collision, at the interaction point, of the two resulting beams, each consisting of electrons, positrons, and photons. The simulations are performed with the code {\sc cain} \cite{CAIN}. This legacy particle physics code can model beam-beam interactions in the strong-field QED regime \cite{gonoskov.rmp.2022}, taking beamstrahlung and pair production into account. The conversion point is handled by a separate part of the code, which uses the linear Compton and Breit-Wheeler processes, together with radiative corrections and higher-order processes, to simulate the production of photons and $e^+e^-$ pairs in the interaction of a particle beam with a laser pulse. These simulations yield the luminosity spectra for $\gamma\gamma$, $\gamma e$, $e^+e^-$, and $e^-e^-$ collisions, allowing us to optimize the laser parameters for the 10 TeV pCM $\gamma\gamma$-collider concept.

\subsection{Electron bunch and laser parameters used in the {\sc cain} simulations}

The following parameters are used in Secs.~III\,B--III\,D. The electron beams are focused at IP and have an energy of 5 TeV, a normalized emittance of 50 nm, a beta function of 0.6 mm at the interaction point, and an energy spread of $\Delta E_0/E_0 = 0.005$. Each bunch contains $N_e = 2.12\times10^9$ particles and has a length of $\sigma_z = 5$ $\mu$m; at 47 kHz the geometric luminosity is $L_{e^-e^-}^{geo}=5.5\times 10^{35}$~cm$^{-2}$s$^{-1}$. The effects of using longer electron bunches are discussed in Appendix II, which connects the present study with established collider designs. In Secs.~III\,E and III\,F electron beams with a 1~mm beta function are used instead, giving $L_{e^-e^-}^{geo}=3.3\times 10^{35}$~cm$^{-2}$s$^{-1}$. These numbers are scaled from a previous wakefield collider design concept \cite{schroeder.jinst.2023} and provide a reasonable starting point for the analysis of the present $\gamma\gamma$-collider case. Unless stated otherwise, the laser pulses are circularly polarized and are focused to a spot size of $20\lambda$ at the conversion point, with a wavelength ranging from 2~$\mu$m to 0.25~nm (laser photon energy from $\hbar\omega_0=0.65$~eV to 5~keV) and a duration ranging from 60~fs to 40~ps. These parameters were chosen to ensure that the optimal $\gamma\gamma$-collider operating point lies inside the parameter space, and that the parameter space is broad enough to identify the physical effects governing the optimization. The dimensionless field amplitude of the circularly polarized laser, $a_0=0.6\,(I[10^{18}~\mbox{W/cm}^2]\,\lambda_0^2[\mu\mbox{m}])^{1/2}$, is set to $a_0=0.3$ for all wavelengths considered in this paper unless noted otherwise; the dependence of the $\gamma\gamma$-collider performance on $a_0$ is addressed later in this study. This value was chosen to keep the interaction in the weakly nonlinear regime during the conversion stage. Since the linear Compton and Breit-Wheeler processes depend significantly on the laser and electron beam polarizations, we choose $2\lambda_e P_c=-0.9$ in order to maximize the high-energy photon production (see Ref.~\cite{tesla.ijmpa.2004} and Appendix I for details), where $P_c$ is the mean laser photon helicity and $\lambda_e$ is the mean electron helicity. A short overview of the {\sc cain} code, together with a summary of the electron beam and laser parameters used in the simulations throughout the paper, is given in Appendix IV. 

\subsection{Conversion stage simulations: benchmarking the 1D model against {\sc cain}}

\begin{figure}[!ht]
\centering
\includegraphics[width=8.6cm]{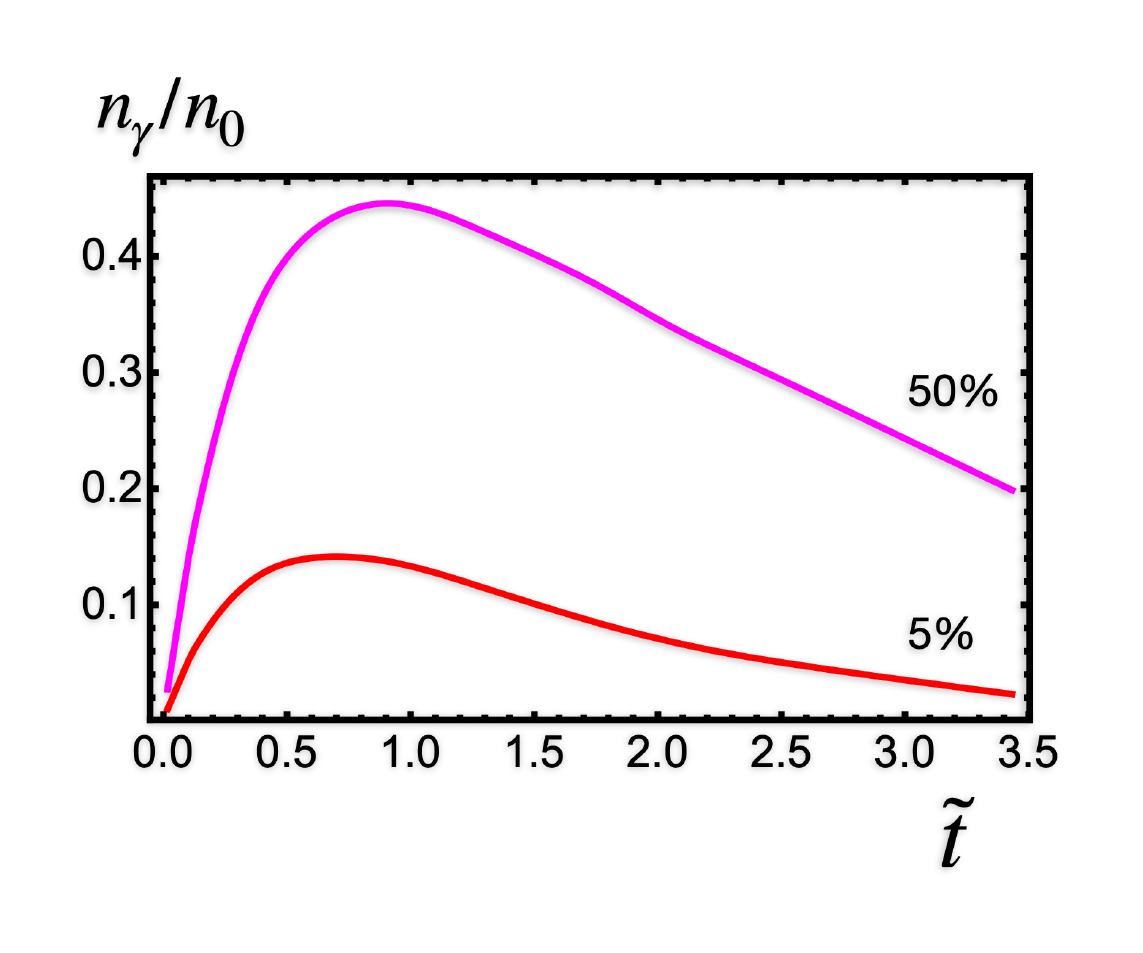}
\caption{Number of high-energy photons from {\sc cain} simulations, normalized to the number of electrons in the beam, as a function of the laser pulse duration. The pulse duration is given in units of the electron mean lifetime with respect to the Compton process. The red curve shows the number of photons with energy $\hbar\omega^\prime>0.95E_0$ and the magenta curve the number with $\hbar\omega^\prime>0.5E_0$. Here $x=191$, $a_0=0.3$, $\hbar\omega_0=2.5$~eV, and $E_0=5$~TeV.}
\label{fig:number of photons_CAIN}
\end{figure}

We first benchmark the 1D analytical model against the {\sc cain} results. Here we simulate only the conversion stage, i.e., the interaction of an electron beam with a scattering laser. The electron beam parameters are the same as above, but the laser pulse length was chosen equal to the Rayleigh length, in order to make the interaction quasi-1D. It was varied from 300~fs to 100~ps at fixed $a_0=0.3$, which implies a corresponding variation of the focal spot radius. Figure~\ref{fig:number of photons_CAIN} shows the resulting dependence of the number of photons on the laser duration for two characteristic photon populations, one with energy exceeding $0.95E_0$ and one with energy exceeding $0.5E_0$. Both curves have a shape similar to the 1D result, but their maxima differ from it, with $n^{max}_\gamma(>0.95 E_0)=0.14n_0$ and $n^{max}_\gamma(>0.5 E_0)=0.45n_0$. If we account for the fact that at $x=191$ the emission rate of photons with energy $>0.95E_0$ is $0.41W_C$, the 1D model gives the estimate $n^{max}_\gamma(>0.95 E_0)=0.41n^{max}_\gamma\approx 0.12n_0$, in good agreement with the {\sc cain} result. The discrepancy in the maximum number of photons with energy $>0.5E_0$ arises because the 1D model omits pairs of successive emissions, for example, in which the first photon has energy $<0.5E_0$ and the second has energy $>0.5E_0$; the model therefore underestimates the number of high-energy photons. From the analytical model we expect the high-energy $\gamma\gamma$ luminosity to be approximately one order of magnitude smaller than the $e^-e^-$ geometric luminosity, and to be very sensitive to the distance between the conversion point and the interaction point. The estimates place this distance at the millimeter scale, which is of the order of the laser pulse length that maximizes the number of photons surviving to the end of the conversion stage. 

\subsection{$\gamma\gamma$-collider simulation results for scattering lasers of different wavelength}

\begin{figure*}[!ht]
\centering
\includegraphics[width=7.5cm]{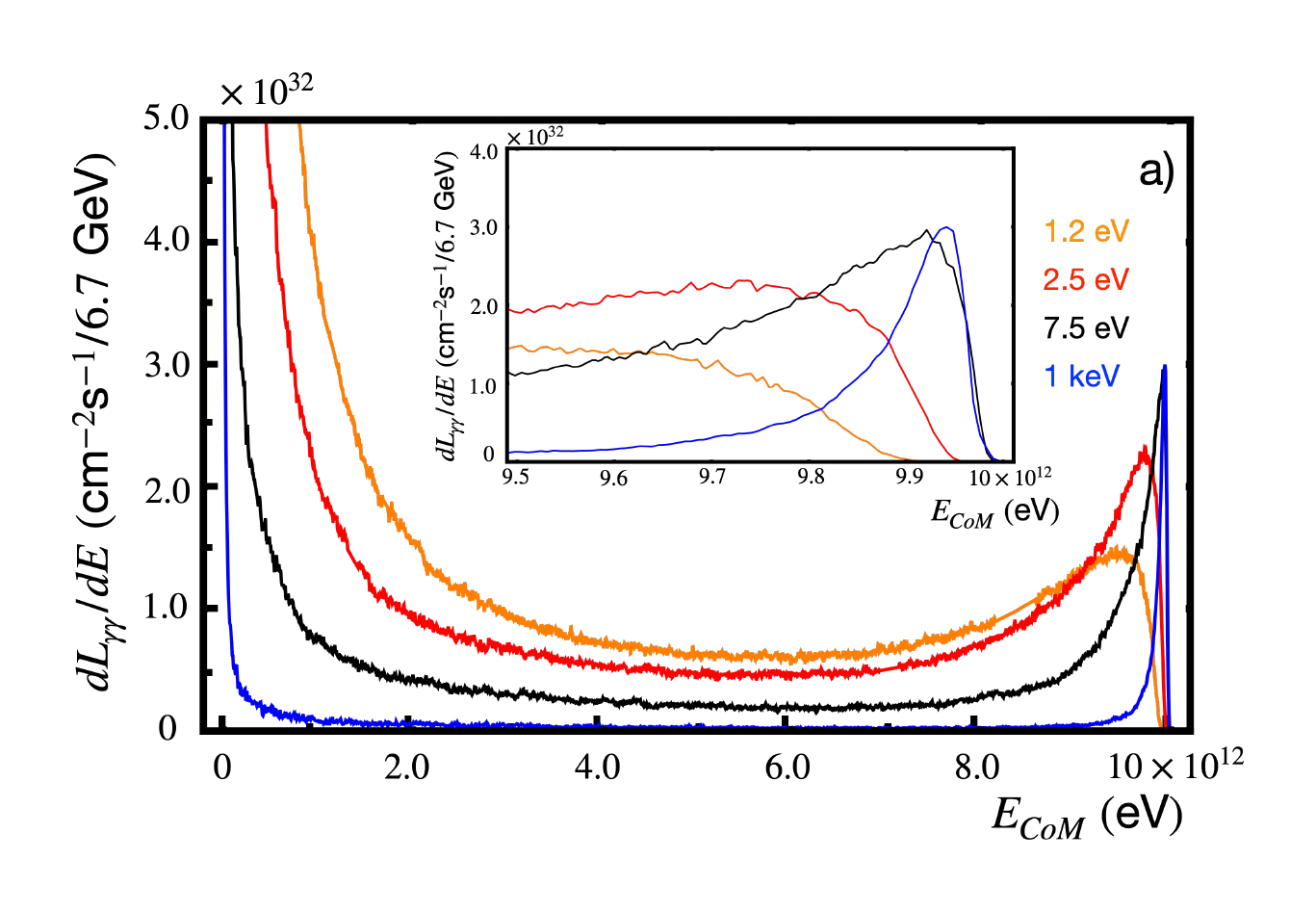}
\includegraphics[width=7.5cm]{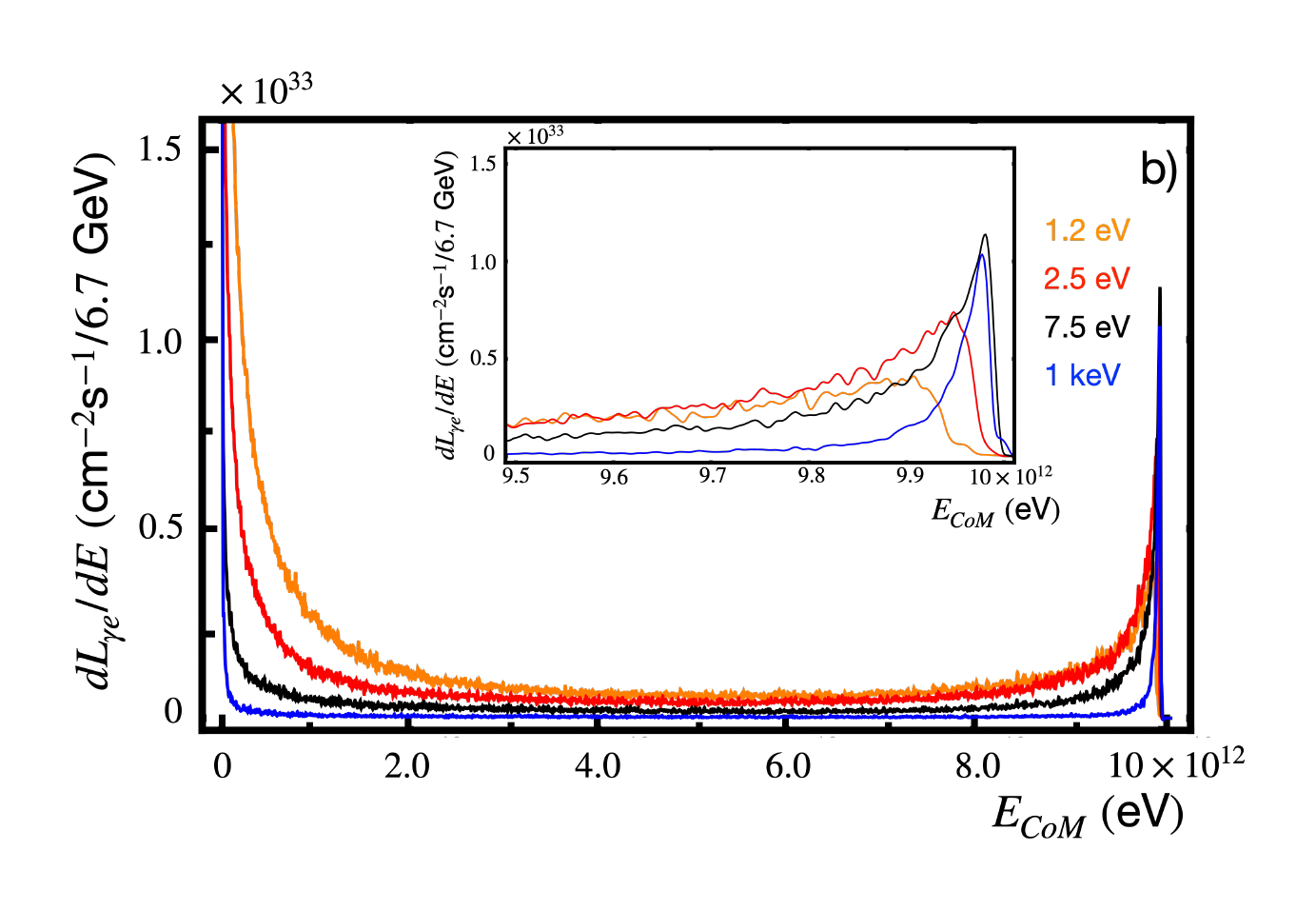}
\includegraphics[width=7.5cm]{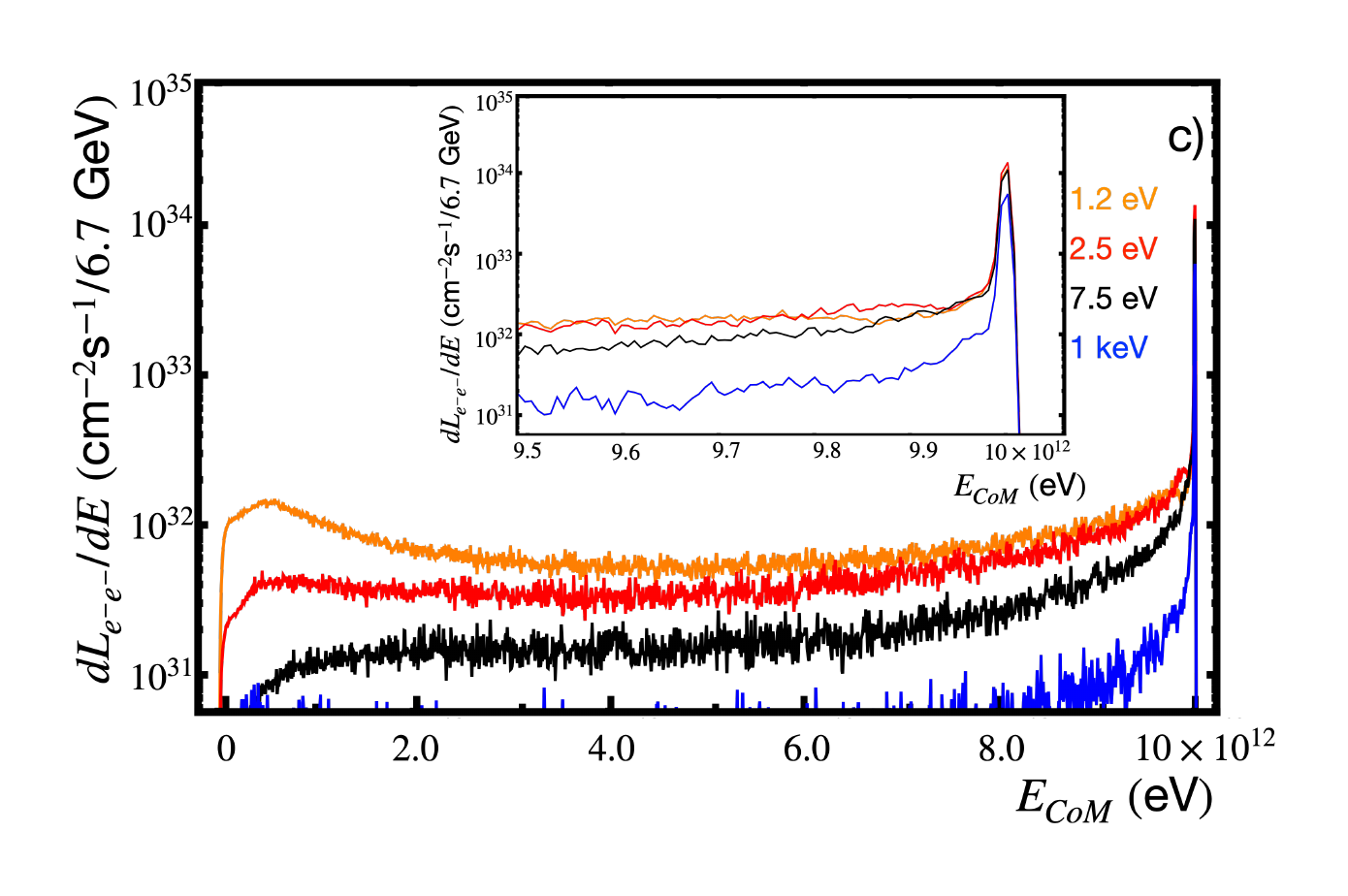}
\includegraphics[width=7.5cm]{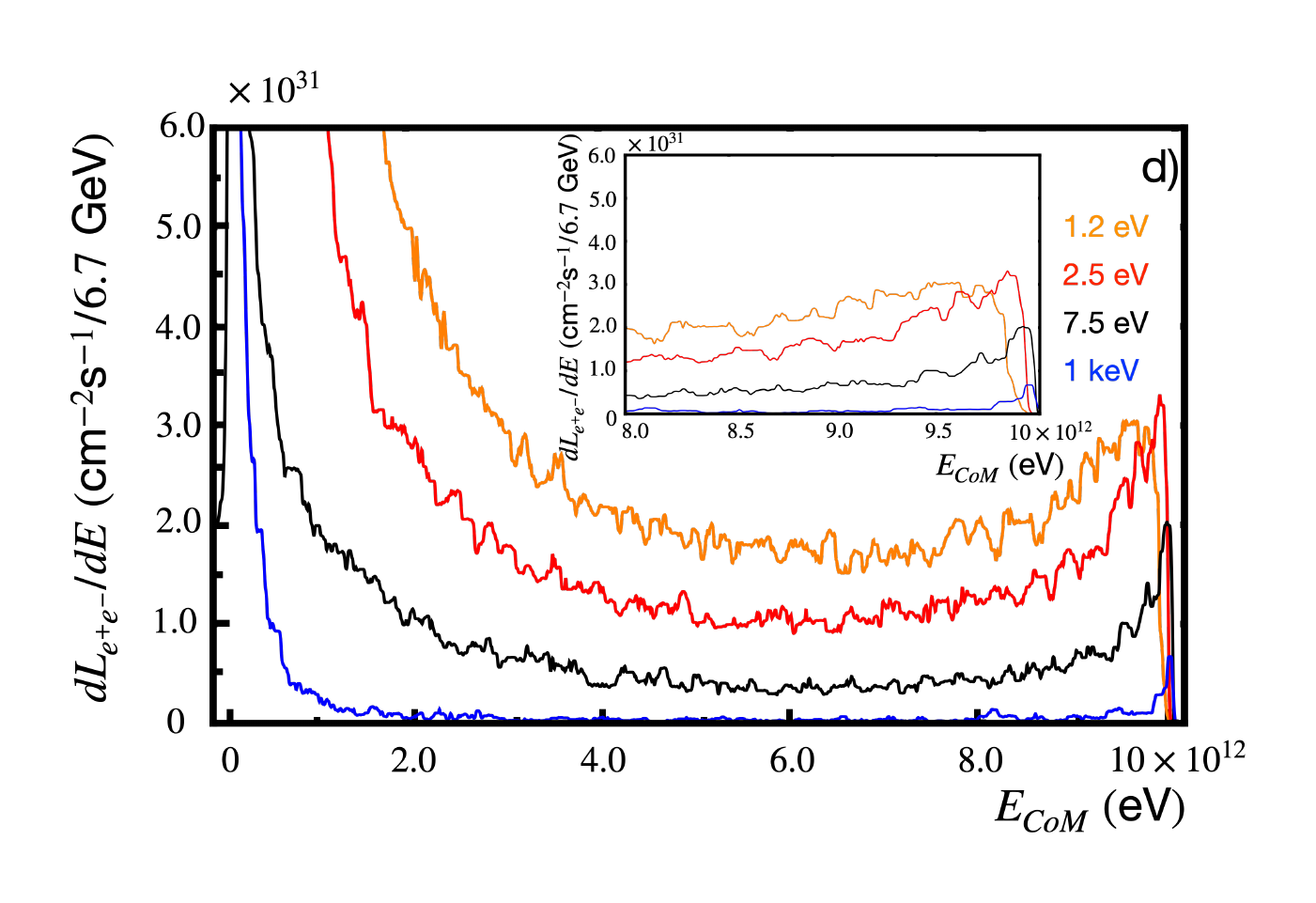}
\caption{The $\gamma\gamma$ (a), $\gamma e$ (b), $e^-e^-$ (c), and $e^+e^-$ (d) luminosity spectra for different scattering laser frequencies: $\hbar\omega_0=1.2$~eV, $c\tau=8$~mm (orange); 2.5~eV, $c\tau=6$~mm (red); 7.5~eV, $c\tau=5.3$~mm (black); and 1~keV, $c\tau=80$~$\mu$m (blue). The insets show the luminosity spectra near the 10~TeV peak.}
\label{fig:lumi}
\end{figure*}

As a first step in analyzing the possible performance of a $\gamma\gamma$-collider, we model the conversion and interaction stages using the electron beam and laser parameters defined above for four values of the laser photon energy, $\hbar\omega_0=1.2$, 2.5, and 7.5~eV and 1~keV, with laser pulse lengths of 8~mm, 6~mm, 5.3~mm, and 80~$\mu$m, respectively. The distance between the conversion and interaction points is chosen to match Eq.~(\ref{dcp}) approximately, and the laser pulse length is set to twice this distance, so as to maximize the number of photons without unduly increasing the radius of the photon beam at the interaction point. Typical $\gamma\gamma$, $\gamma e^-$, and $e^-e^-$ luminosity spectra for different scattering laser wavelengths are shown in Fig.~\ref{fig:lumi}. As the parameter $x$ increases from 92 at $\hbar\omega_0=1.2$~eV to $7.7\times10^4$ at $\hbar\omega_0=1$~keV, the $\gamma\gamma$ luminosity spectra develop a sharp peak at the maximum center-of-mass energy, the spectrum at 1~keV exhibiting a width of 75~GeV FWHM, or $\Delta E_{CoM}/E_{CoM}=7.5\times 10^{-3}$ (compared with the initial $\Delta E_0/E_0=5\times 10^{-3}$). The peaks produced by near-optical laser pulses, which we define as those with photon energies from $\hbar\omega_0=0.65$~eV to 10~eV, are an order of magnitude wider, ranging from $\Delta E_{CoM}/E_{CoM}=2.5\times 10^{-2}$ at $\hbar\omega_0=7.5$~eV to $2.3\times 10^{-1}$ at $\hbar\omega_0=1.2$~eV, although their maximum values are similar (see Fig.~\ref{fig:lumi}a and Table~\ref{tab:gg lumi}). The $\gamma e^-$ and $e^-e^-$ luminosity spectra also exhibit sharp peaks near the maximum center-of-mass energy, and near-optical laser pulses lead to higher total luminosities in both cases (see Figs.~\ref{fig:lumi}b and \ref{fig:lumi}c). The Breit-Wheeler process during the conversion stage gives rise to a population of $e^+e^-$ pairs in addition to those produced in the beam-beam interaction, and all of these collide at the interaction point; the corresponding luminosity spectra are shown in Fig.~\ref{fig:lumi}d. All of these spectra are characterized by an enhanced number of high-energy pairs, although the higher the laser frequency, the smaller the corresponding total luminosity, which can be explained in part by the shorter $d_{CP}$. Both near-optical and X-ray lasers thus enable $\gamma\gamma$-collider operation even in the presence of prolific $e^+e^-$ pair production. The photon distributions obtained with near-optical lasers lead to more $\gamma\gamma$ collisions at almost every energy, with potential benefits for new particle searches \cite{chigusa.arxiv.2025,cipressi.arxiv.2026,fraser.arxiv.2026}. 

\begin{table}
\caption{\label{tab:gg lumi} Parameters of the high-energy $\gamma\gamma$ luminosity peak in Fig.~\ref{fig:lumi}a.}
\begin{tabular}{c|c||c|c|c|}
\hline
$\hbar\omega_0$ & $c\tau$ & $max[dL_{\gamma\gamma}/dE]\times 10^{-32}$ & $E_{CoM}$ & $\Delta E_{CoM}$\\ \hline\hline
1.2 eV & 8 mm & $1.5$ cm$^{-2}$ $s^{-1}$/6.7 GeV & 9.5 TeV & 2.2 TeV  \\
2.5 eV & 6 mm & $2.3$ cm$^{-2}$ $s^{-1}$/6.7 GeV & 9.7 TeV & 0.7 TeV \\
7.5 eV & 5.3 mm & $3.0$ cm$^{-2}$ $s^{-1}$/6.7 GeV & 9.925 TeV & 0.25 TeV \\
1.0 keV & 80 $\mu$m & $3.0$ cm$^{-2}$ $s^{-1}$/6.7 GeV & 9.95 TeV & 75 GeV \\
\hline
\end{tabular}
\end{table}

\subsection{Optimization of the $\gamma\gamma$ luminosity through scattering laser wavelength and duration selection}

\begin{figure}[!ht]
\centering
\includegraphics[width=8cm]{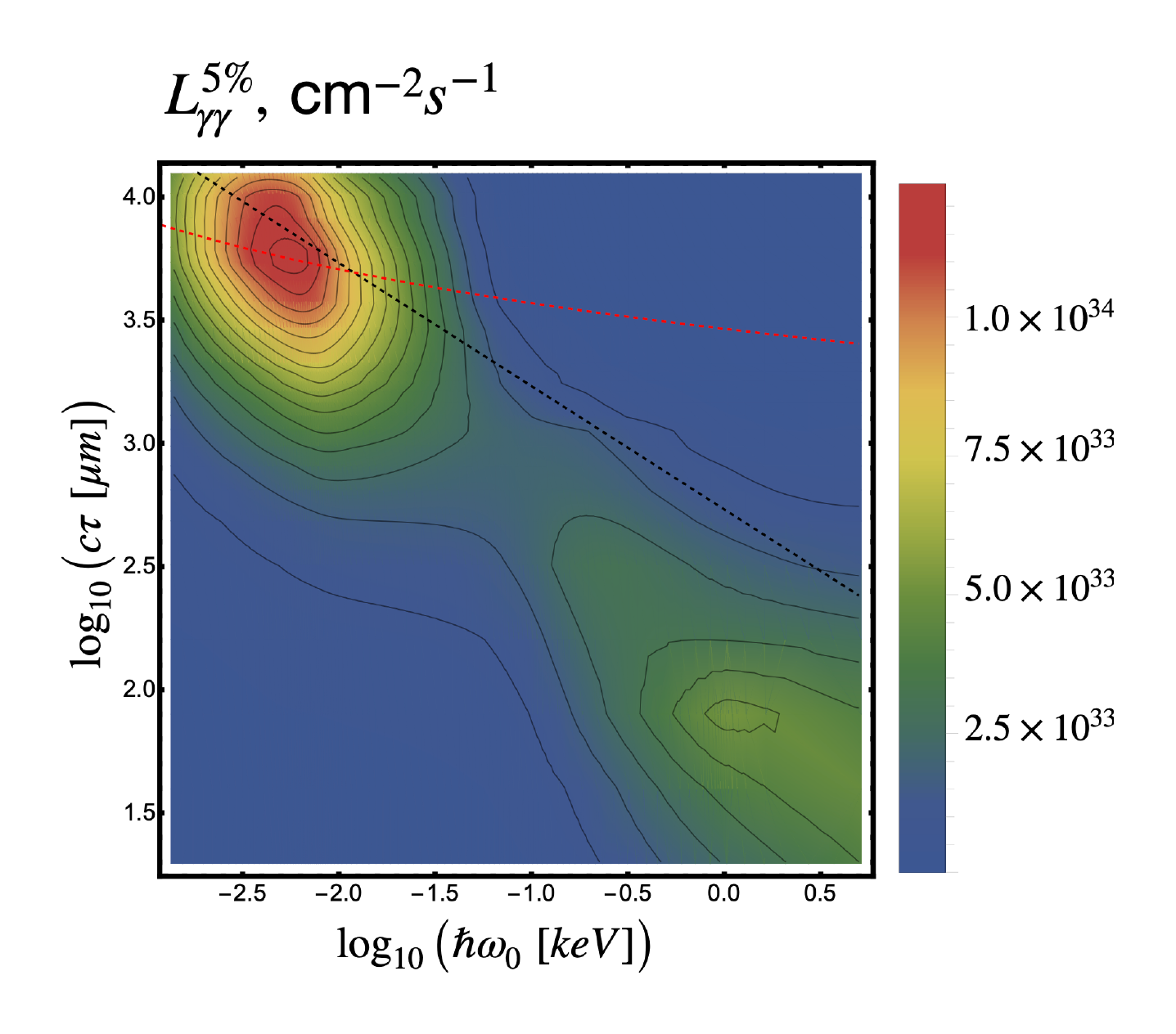}
\includegraphics[width=8cm]{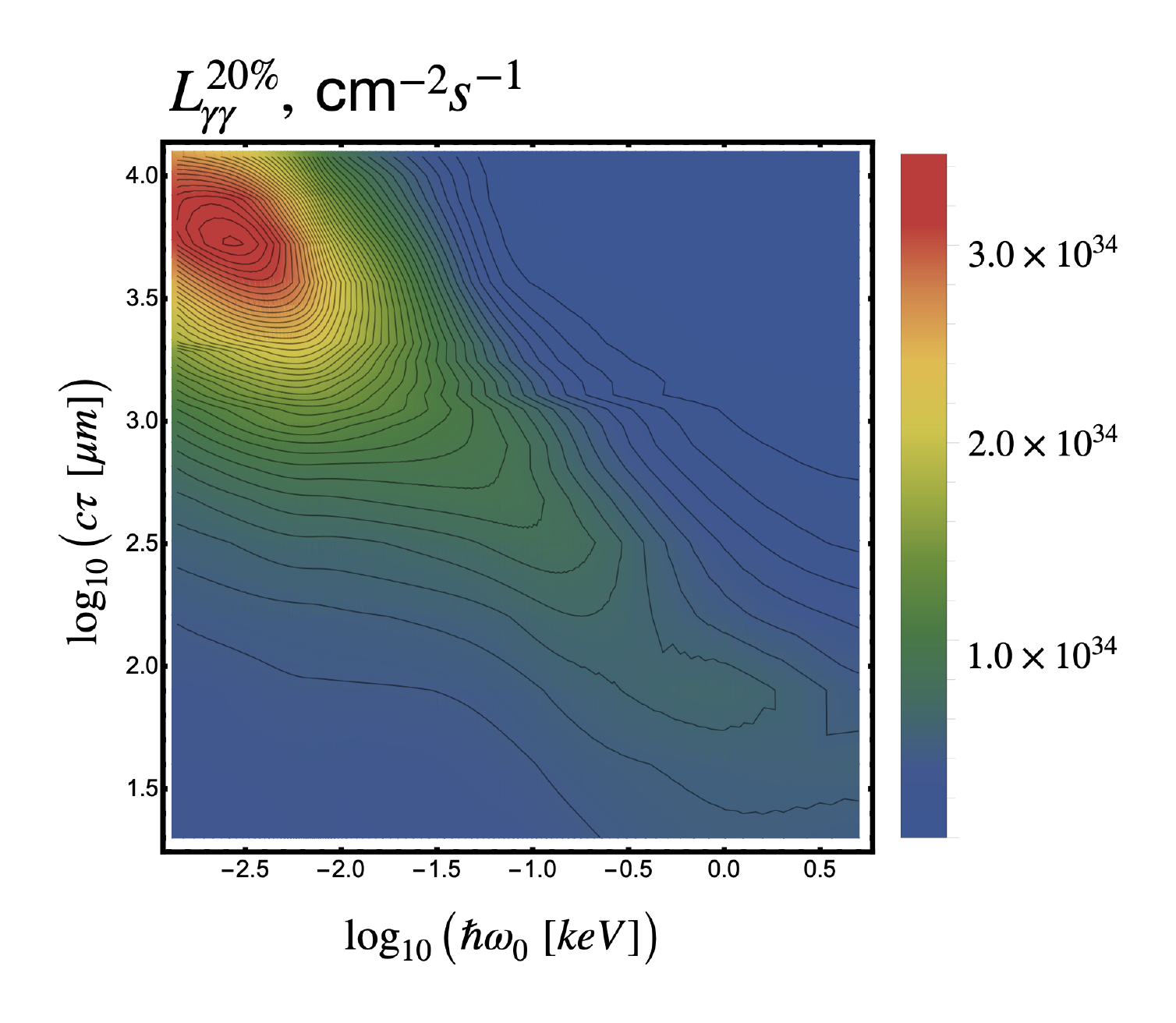}
\caption{(a) Dependence of the 5\% luminosity on laser frequency and laser duration. The dashed black curve is given by $c\tau=2\sqrt{\epsilon\beta}/\theta_b$ (with $b=0.95$) and the dashed red line by Eq.~(\ref{eq:tmax}). (b) Dependence of the 20\% luminosity on laser frequency and laser duration.}
\label{fig:lumi scan}
\end{figure}

To characterize the search potential of a $\gamma\gamma$-collider, we studied the dependence of the partial $\gamma\gamma$ luminosities on the scattering laser frequency and duration (see Fig.~\ref{fig:lumi scan}), taking the distance between the conversion and interaction points to be half the laser pulse length. The partial luminosities are defined as the luminosities of collisions with center-of-mass energies exceeding a given value, i.e., $L_{\gamma\gamma}^{5\%}=L_{\gamma\gamma}(E_{CoM}>0.95\times 2E_0)$ and $L_{\gamma\gamma}^{20\%}=L_{\gamma\gamma}(E_{CoM}>0.8\times 2E_0)$. For $L_{\gamma\gamma}^{5\%}=L_{\gamma\gamma}^{5\%}(\hbar\omega_0,c\tau_0)$, two well-defined maxima are apparent: one at $\hbar\omega_0=5$~eV, $c\tau=6$~mm, and $W_L=125$~J, and the other at $\hbar\omega_0=1$~keV, $c\tau=80$~$\mu$m, and $W_L=1.7$~J. The near-optical laser yields $L_{\gamma\gamma}^{5\%}=1.3\times10^{34}$~cm$^{-2}$s$^{-1}$, whereas the X-ray laser yields $5.3\times10^{33}$~cm$^{-2}$s$^{-1}$ (the total and partial 5\% luminosities for these two maxima are listed in Table~\ref{tab:compare 5 eV and 1 keV}). If a broader range of center-of-mass energies is chosen ($E_{CoM}>0.8\times 2E_0$), the maximum at keV laser photon energies disappears, while the near-optical maximum shifts towards lower photon energies ($\hbar\omega_0=2.5$~eV), reaching $L_{\gamma\gamma}^{20\%}=3.7\times10^{34}$~cm$^{-2}$s$^{-1}$. The maximum value of $L_{\gamma\gamma}^{20\%}$ at 1~keV is $6.4\times 10^{33}$~cm$^{-2}$s$^{-1}$, nearly six times smaller.

We note that the region of parameter space with low frequencies and short durations does not yield high partial luminosities, which can be explained by the insufficient number of high-energy photons produced there. The region with high frequencies and long durations likewise yields low partial luminosities, but for a different reason: longer laser pulses require a longer $d_{CP}$, which results in a wider photon beam at the interaction point, especially at high laser frequencies, and hence in a lower luminosity. The partial luminosity scans show that the optimization of a $\gamma\gamma$-collider, which we take to be the maximization of the partial luminosity, approximately follows the condition $d_{CP}=\sqrt{\epsilon\beta}/\theta_b$, with the maximum at $\hbar\omega_0=5$~eV and $c\tau=6$~mm lying very close to the 1D analytical result of $\hbar\omega_0=10$~eV and $c\tau=5$~mm given above (see Fig.~\ref{fig:lumi scan}a). The existence of the second maximum, at $\hbar\omega_0=1$~keV and $c\tau=80$~$\mu$m, can be ascribed to the enhancement of high-energy photon production at large $x$ and to polarization effects. We note that there is a potential alternative direction for $\gamma\gamma$-collider optimization: increasing the interaction length without significantly increasing the scattering laser energy. As noted in Ref.~\cite{telnov.nima.2001b}, a traveling laser focus can relax the limitations posed by the Rayleigh length. Such a traveling focus can be realized with the flying-focus technique \cite{froula.natphot.2018}, and its application to $\gamma\gamma$-collider optimization should be explored in detail in future work.

\begin{table}[]
\caption{\label{tab:compare 5 eV and 1 keV} Total and partial 5\% luminosities, in units of cm$^{-2}$s$^{-1}$, for the 5~eV and 1~keV $\gamma\gamma$-colliders corresponding to the maxima in Fig.~\ref{fig:lumi scan}a.}
\begin{tabular}{c||c|c|c|c|}
\multirow{2}{*}{} & \multicolumn{2}{l|}{$\hbar\omega_0=5$ eV} & \multicolumn{2}{l|}{$\hbar\omega_0=1$ keV} \\ \hline
 &      total     &    5\%      &     total      &   5\%       \\ \hline
$L_{\gamma\gamma}$ &    $1.2\times 10^{35}$       &    $1.3\times 10^{34}$      &    $1.8\times 10^{34}$       &    $5.3\times 10^{33}$      \\
$L_{\gamma e^-}$&     $8.6\times 10^{34}$      &     $2.0\times 10^{34}$     &    $1.8\times 10^{34}$       &    $8.1\times 10^{33}$      \\
$L_{e^-e^-}$ &      $5.9\times 10^{34}$     &     $2.8\times 10^{34}$     &     $1.8\times 10^{34}$      &      $1.2\times 10^{34}$    \\
$L_{e^+e^-}$ &    $2.0\times 10^{34}$       &    $1.3\times 10^{33}$      &   $9.1\times 10^{33}$        &   $2.0\times 10^{32}$  \\ \hline    
\end{tabular}
\end{table}

\subsection{Comparison of different operational scenarios ($e^+e^-$, $e^-e^-$, and $\gamma\gamma$) of a 10 TeV collider}

\begin{figure*}[!ht]
\centering
\includegraphics[width=8cm]{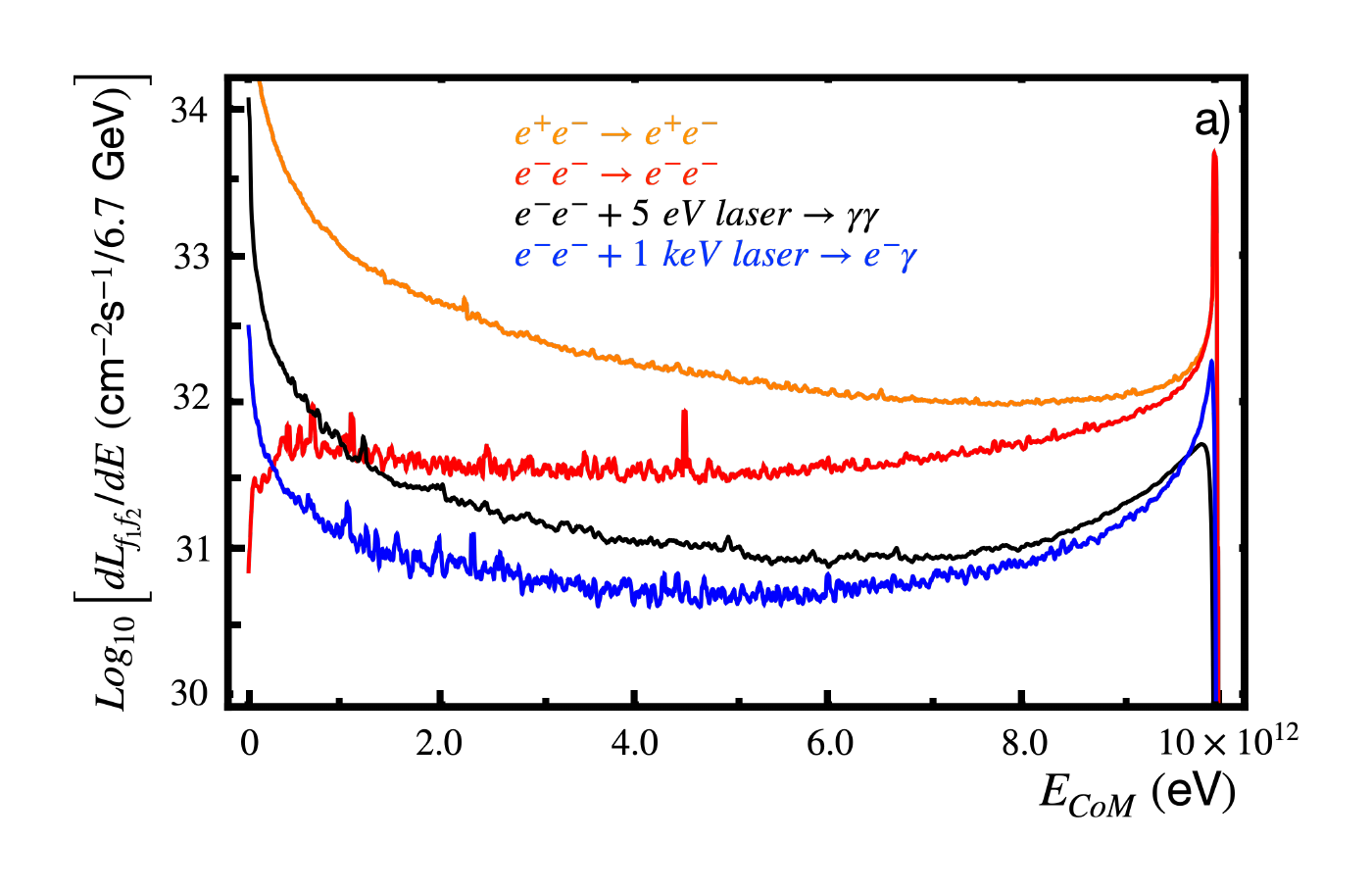}
\includegraphics[width=8cm]{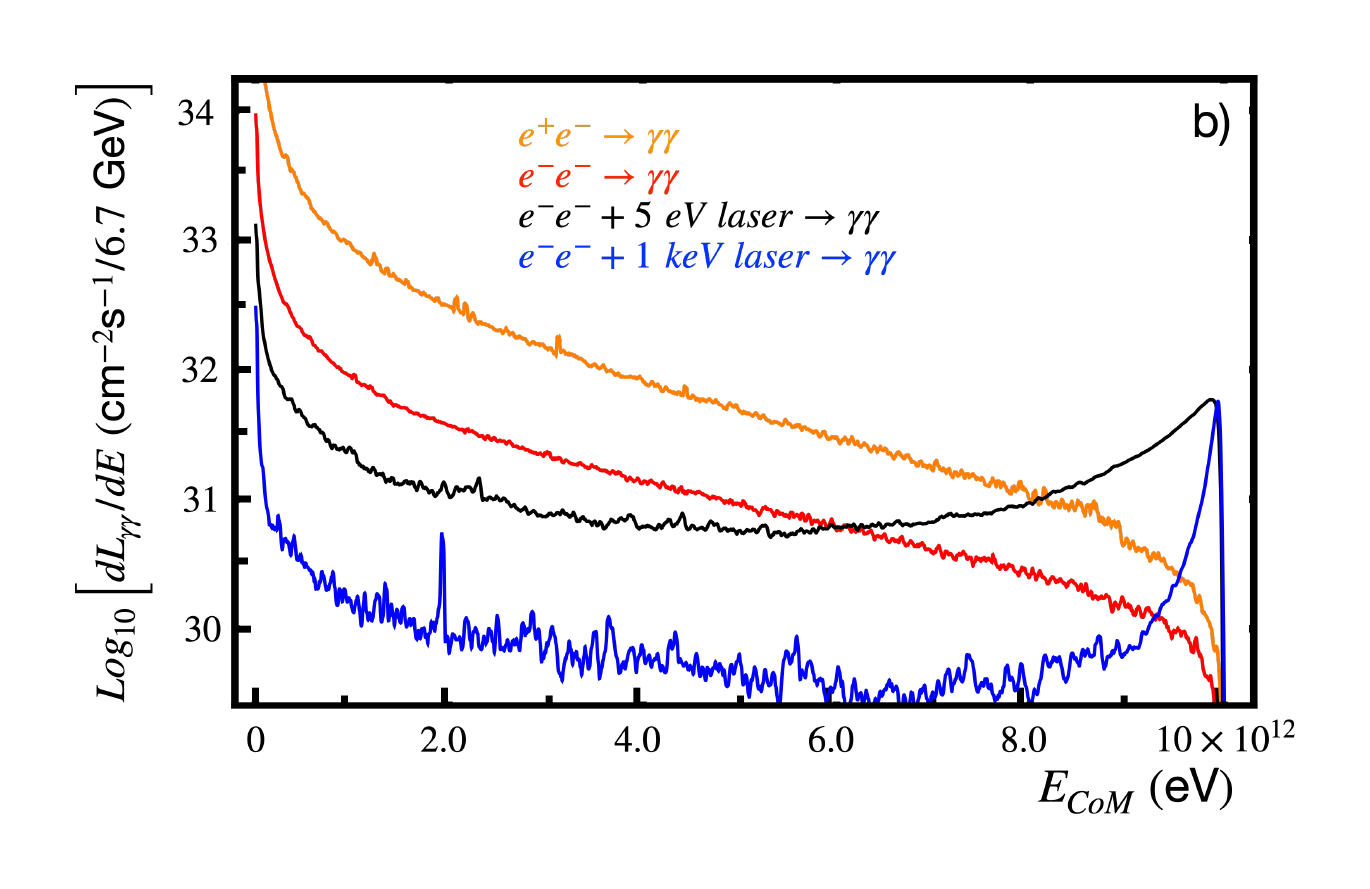}
\includegraphics[width=8cm]{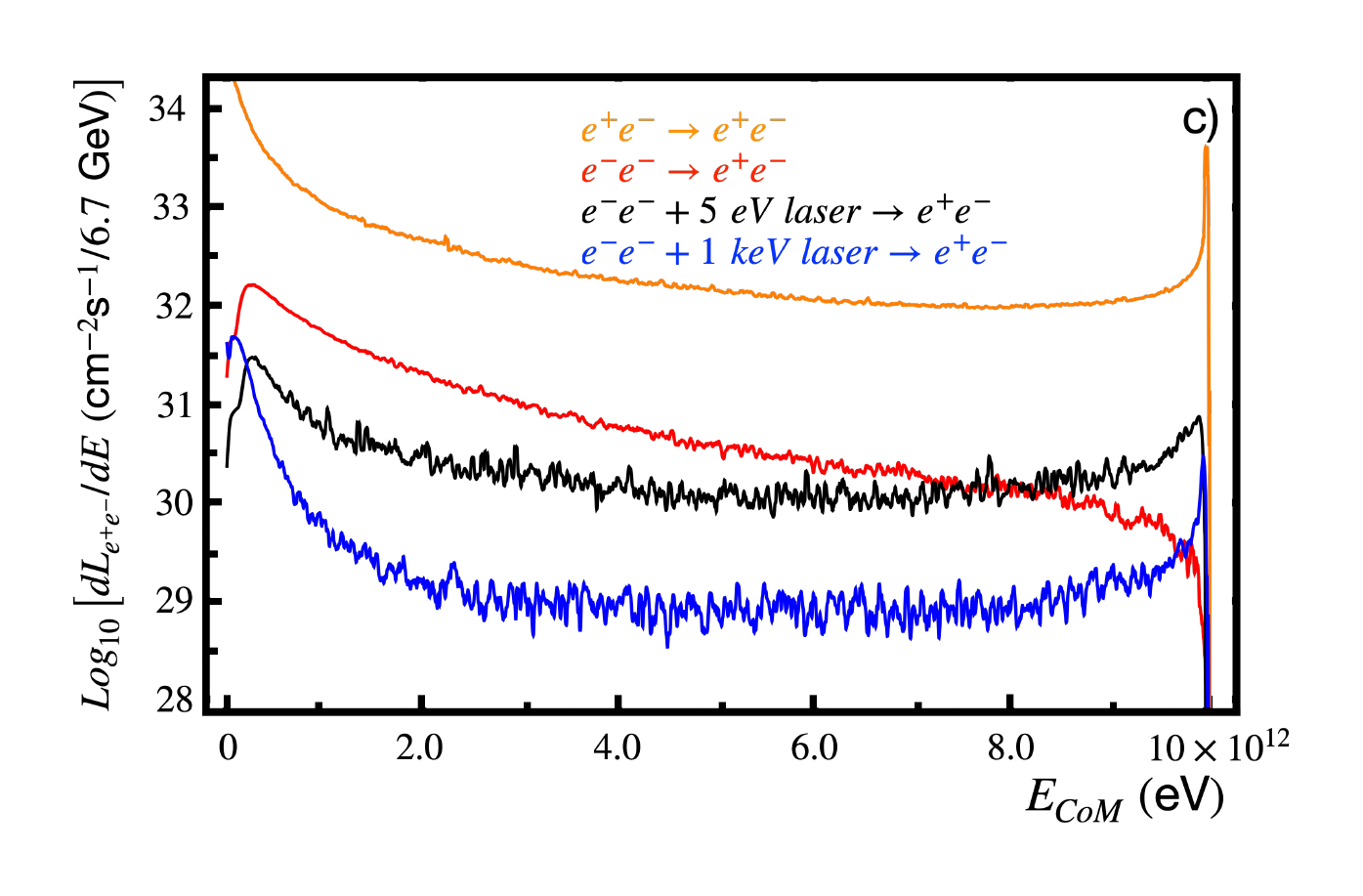}
\includegraphics[width=8cm]{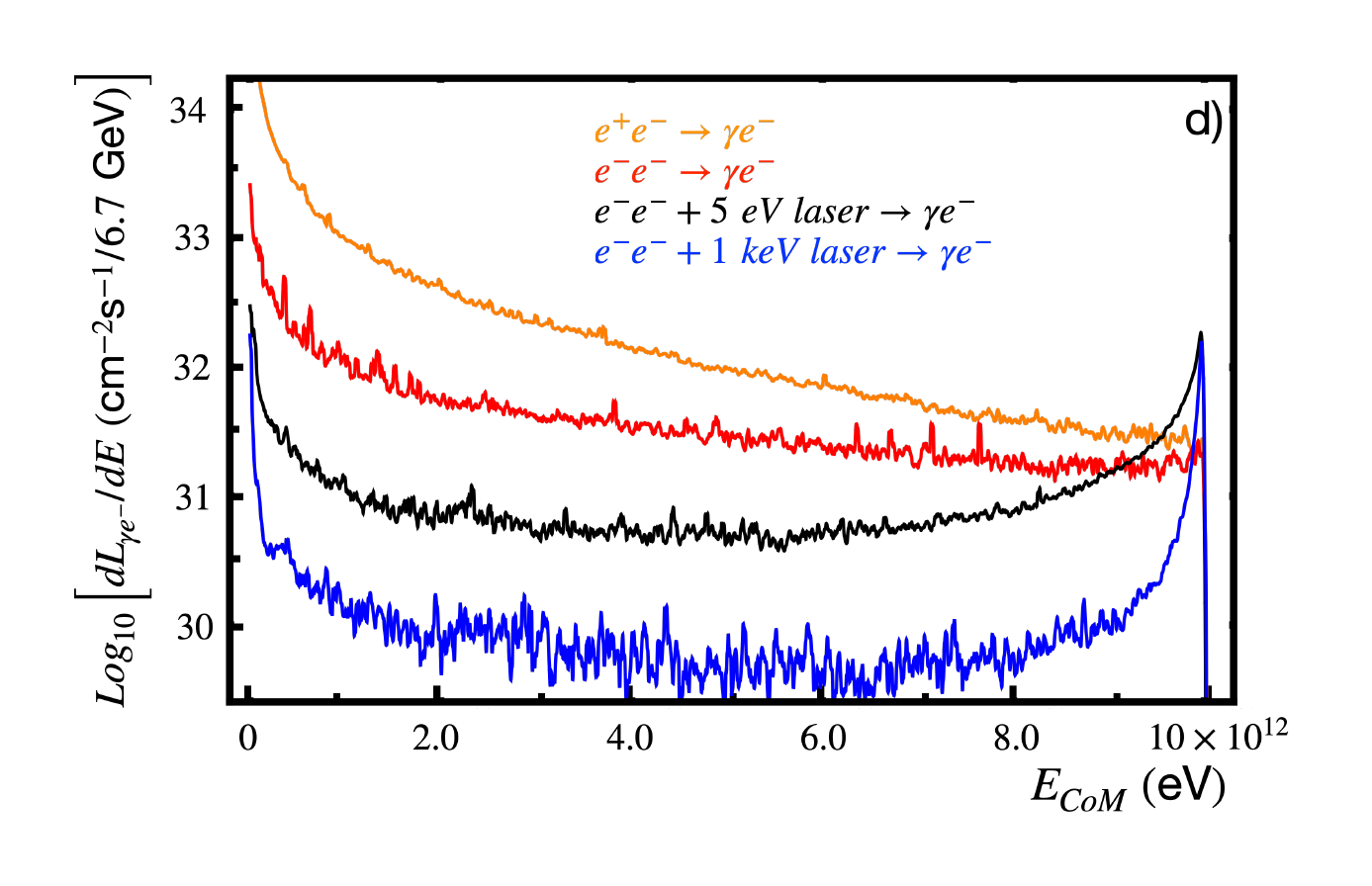}
\caption{(a) Luminosity spectra for different collider and collision options: $e^+e^-$ accelerator with $e^+e^-$ collisions at the IP ($f_1=e^+$, $f_2=e^-$, orange curve), $e^-e^-$ accelerator with $e^-e^-$ collisions at the IP ($f_1=e^-$, $f_2=e^-$, red curve), $e^-e^-$ accelerator with $\gamma\gamma$ collisions generated by a 5~eV laser ($f_1=\gamma$, $f_2=\gamma$, black curve), and $e^-e^-$ accelerator with $\gamma e^-$ collisions generated by a 1~keV laser ($f_1=\gamma$, $f_2=e^-$, blue curve). (b) $\gamma\gamma$, (c) $e^+e^-$, and (d) $\gamma e^-$ collision luminosity spectra for the different collider options: $e^+e^-$ accelerator (orange curves), $e^-e^-$ accelerator (red curves), $e^-e^-$ accelerator with a 5~eV scattering laser (black curves), and $e^-e^-$ accelerator with a 1~keV scattering laser (blue curves).}
\label{fig:lumi compare}
\end{figure*}

To compare different operational scenarios of a 10 TeV energy-frontier collider, we show in Fig.~\ref{fig:lumi compare} the luminosity spectra for a range of collider and collision options. Here the electron and positron beams have radii of 2.24~nm at the interaction point ($\gamma\epsilon=50$~nm, $\beta=1.0$~mm), giving a geometric luminosity of $L_{ee}^{geo}=3.3\times 10^{35}$~cm$^{-2}$s$^{-1}$; the corresponding total and partial luminosities are listed in Table~\ref{tab:collider lumi}. Figure~\ref{fig:lumi compare} shows the spectra for an $e^+e^-$ accelerator with $e^+e^-$ collisions, an $e^-e^-$ accelerator with $e^-e^-$ collisions, an $e^-e^-$ accelerator with $\gamma\gamma$ collisions (generated by a 5~eV laser), and an $e^-e^-$ accelerator with $\gamma e^-$ collisions (generated by a 1~keV laser) at the IP. These options were chosen to represent the main collision type for each collider, the exception being the $e^-e^-$ accelerator with a 1~keV laser, for which the $\gamma e^-$ spectrum is shown in order to indicate how this channel relates to the others. All four spectra are similar in shape, each having a spike in the high-energy part with a broad extension to lower energies, but their total and partial luminosities differ.

\begin{table*}
\begin{tabular}{c||c|c|c|c|c|c|c|c}
\hline
Collider (collision) options & $L_{f_1 f_2}$ & $L_{f_1 f_2}^{5\%}$ & $L_{\gamma\gamma}$ & $L_{\gamma\gamma}^{5\%}$ & $L_{e^+ e^-}$ & $L_{e^+ e^-}^{5\%}$ & $L_{\gamma e}$ & $L_{\gamma e}^{5\%}$ \\ \hline\hline
$e^+e^-$ ($f_1=e^+$, $f_2=e^-$) & $4.3\times 10^{36}$ & $9.9\times 10^{34}$ & $6.8\times 10^{36}$ & $2.5\times 10^{32}$ & $4.3\times 10^{36}$ & $9.9\times 10^{34}$ & $5.2\times 10^{36}$ & $4.7\times 10^{33}$  \\
$e^-e^-$ ($f_1=e^-$, $f_2=e^-$) & $3.0\times 10^{35}$ & $1.1\times 10^{35}$ & $3.0\times 10^{35}$ & $9.3\times 10^{31}$ & $6.8\times 10^{34}$ & $2.4\times 10^{29}$ & $3.5\times 10^{35}$ & $4.2\times 10^{33}$ \\
5 eV $\gamma\gamma$ ($f_1=\gamma$, $f_2=\gamma$) & $8.9\times 10^{34}$ & $8.4\times 10^{33}$ & $8.9\times 10^{34}$ & $8.4\times 10^{33}$ & $1.4\times 10^{34}$ & $8.5\times 10^{32}$ & $6.1\times 10^{34}$ & $1.3\times 10^{34}$ \\
1 keV $\gamma \gamma$ ($f_1=\gamma$, $f_2=e^-$) & $1.2\times 10^{34}$ & $5.1\times 10^{33}$ & $1.1\times 10^{34}$ & $3.2\times 10^{33}$ & $5.8\times 10^{33}$ & $1.2\times 10^{32}$ & $1.1\times 10^{34}$ & $5.1\times 10^{33}$ \\
\hline
\end{tabular}
\caption{\label{tab:collider lumi} Total and partial luminosities for the different collider options, in units of cm$^{-2}$s$^{-1}$. The geometric luminosity of the $e^+e^-$ and $e^-e^-$ collisions is $L_{geo}^{ee}=3.3\times 10^{35}$~cm$^{-2}$s$^{-1}$.}
\end{table*}

In Fig.~\ref{fig:lumi compare}b we show the $\gamma\gamma$ collisions for four different collider versions: $e^+e^-$, $e^-e^-$, and $\gamma\gamma$ powered by either a 5~eV or a 1~keV scattering laser. Collisions in the $e^+e^-$ and $e^-e^-$ machines produce significant luminosity, but the high-energy part of their spectra is strongly suppressed relative to both the 5~eV and the 1~keV $\gamma\gamma$ options, which exhibit enhanced collision rates at these energies. In the $e^+e^-$ and $e^-e^-$ colliders the photons are produced through beamstrahlung, with a decaying power-law spectrum, whereas in the $\gamma\gamma$ options a significant fraction of the photons is produced in the conversion stage through linear Compton scattering, giving a spectrum that peaks at high energies. At the interaction point, where the electrons of the initial beams collide with the photons and $e^+e^-$ pairs created at the conversion point, additional photons are produced by beamstrahlung, which boosts the low-energy part of the $\gamma\gamma$ luminosity spectrum.

As mentioned above, the secondary $e^+e^-$ pairs may aid particle searches, since their luminosity can exceed that of present-day colliders (see Table~\ref{tab:collider lumi}). The $e^+e^-$ luminosity spectra for the three collider options ($\gamma\gamma$ being represented by two curves, for 5~eV and 1~keV) are shown in Fig.~\ref{fig:lumi compare}c. $\gamma e^-$ collisions at the interaction point offer a further channel for new particle searches, and the corresponding luminosity spectra are shown in Fig.~\ref{fig:lumi compare}d. Both $\gamma\gamma$ cases dominate at the highest energies, whereas the $e^+e^-$ and $e^-e^-$ cases generate significantly more collisions in total. We note that from the point of view of collision options $e^+e^-$ and $e^-e^-$ colliders are dominated by $e^+e^-$ and $e^-e^-$ ones respectively. Whereas both $\gamma\gamma$ colliders demonstrate a mix of different collision channels: $e^+e^-$, $\gamma e^-$, and $\gamma\gamma$. This might be beneficial for new particle searches and needs to be studied in future as a part of a comprehensive science case for a 10 TeV collider (see, e.g., Refs. \cite{chigusa.arxiv.2025,fraser.arxiv.2026,fraser.arxiv.2026,cipressi.arxiv.2026}, where some of these luminosity spectra were used to inform the modeling of new particle searches).  

\subsection{Dependence of $\gamma\gamma$-collider operation on the scattering laser intensity}

\begin{figure}[!ht]
\centering
\includegraphics[width=8cm]{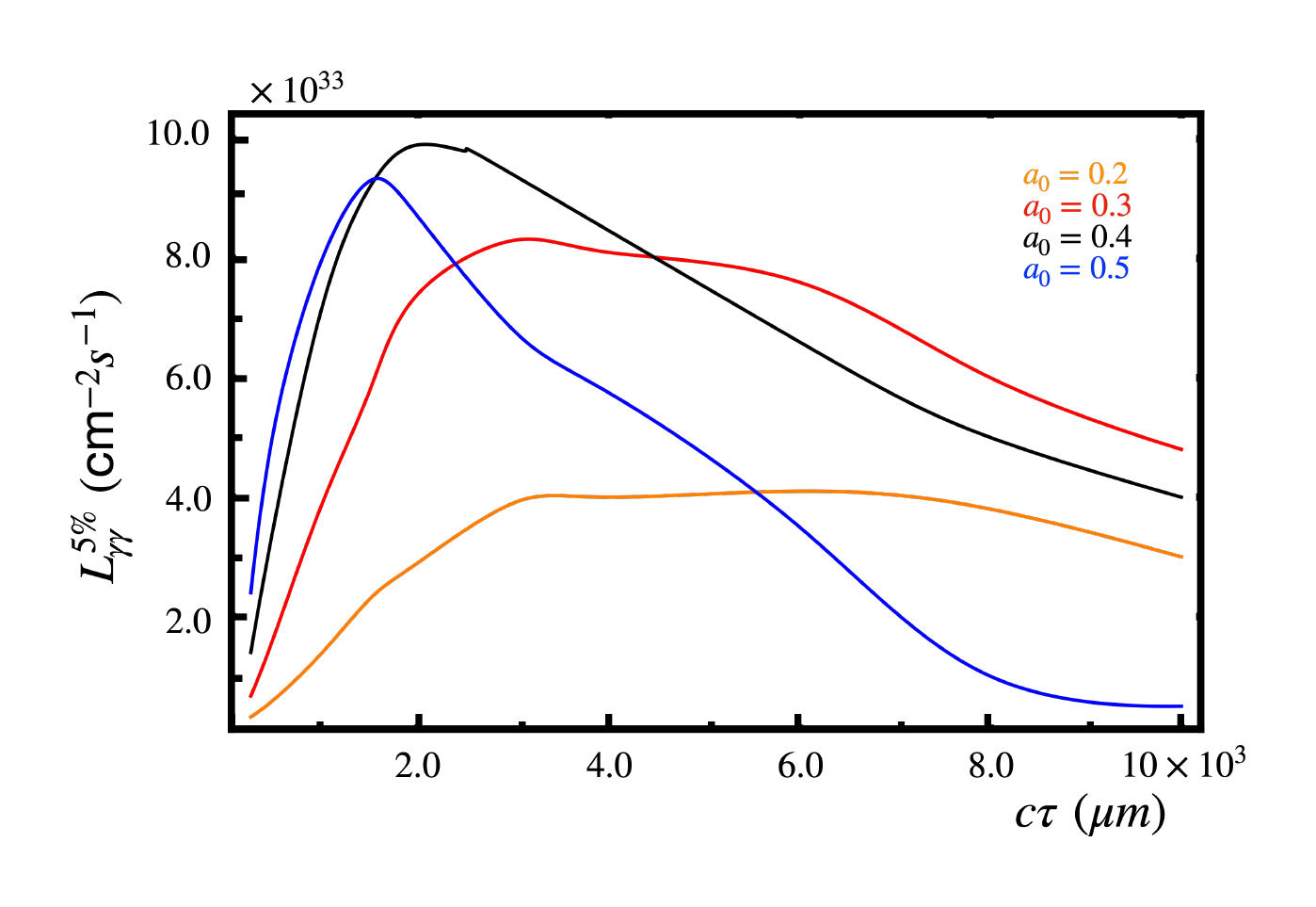}
\caption{Dependence of the partial 5\% $\gamma\gamma$ luminosity on the laser duration for $a_0=0.2$ (orange), 0.3 (red), 0.4 (black), and 0.5 (blue). The other parameters are the same as in Fig.~\ref{fig:lumi compare} for the 5~eV $\gamma\gamma$-collider.}
\label{fig:lumi a0}
\end{figure}

The results presented above were obtained for $a_0=0.3$, which we used for all laser wavelengths and durations considered. Since $a_0$ is an important parameter characterizing the laser-electron interaction, its effect on the collider performance needs to be studied. For the maximum number of generated photons our 1D model yields $n_\gamma^{max}=n_0\kappa^{\kappa/(1-\kappa)}$, which, with the chosen approximations for $W_C$ and $W_{BW}$, is independent of $a_0$. The dependence on $a_0$ can therefore enter only through the laser pulse duration required to reach $n_\gamma^{max}$, since the optimal distance between the conversion and interaction points is determined by the laser duration. In what follows we examine whether this simple explanation holds in the {\sc cain} simulations. To this end, we carried out $\gamma\gamma$-collider simulations with the parameters used in Fig.~\ref{fig:lumi compare} for the 5~eV laser case, taking $a_0=0.2$, 0.3, 0.4, and 0.5. The resulting dependence of the partial 5\% $\gamma\gamma$ luminosity on the laser duration is shown in Fig.~\ref{fig:lumi a0}. The values $a_0=0.3$, 0.4, and 0.5 yield comparable maxima of $L_{\gamma\gamma}^{5\%}$, although the reduced maximum at $a_0=0.5$ can be attributed to nonlinear effects in Compton scattering, which point towards efficient collider operation at moderate $a_0$. For $a_0=0.2$, the effect of the laser duration, i.e., the luminosity degradation caused by the increased distance between the conversion and interaction points, starts to become relevant. Thus, for $\gamma\gamma$-collider designs using relatively high values of $a_0$, the precise value can be varied to optimize energy, cost, and efficiency considerations, whereas lower values of $a_0$ give smaller $\gamma\gamma$ luminosities and may require the optimization of other parameters to mitigate this reduction.       

\section{Conclusions}

In this paper we have reported on the study and optimization of a 10 TeV $\gamma\gamma$-collider concept in which high-energy photon beams are produced by Compton scattering of 5 TeV electron beams off moderate-intensity lasers just before the collider interaction point. Most previous design studies stipulated that the emitted photons should not convert into $e^+e^-$ pairs while propagating inside the laser pulse, which places a lower limit on the laser wavelength that can be used. While for $\gamma\gamma$-collider Higgs factory designs this limit leaves the laser wavelength comfortably in the optical range, any multi-TeV $\gamma\gamma$-collider would require long-wavelength lasers, whose use would pose a significant technological challenge. 

Here we explored a range of laser wavelengths, from optical to X-ray, from the point of view of high-energy photon beam generation in the high-$x$ regime, $x\gg 4.8$. We showed that even in the presence of prolific $e^+e^-$ pair production a significant number of high-energy photons can be generated by adjusting the laser pulse duration so as to balance high-energy photon production by Compton scattering against photon loss to Breit-Wheeler pair creation. Over this wide range of wavelengths, the maximum number of high-energy photons varied from 0.5 to 0.3 per initial electron in the bunch. We also identified the importance of minimizing the distance between the conversion and interaction points, so that the photon beams have the smallest possible spot size at collision, which prevents degradation of the $\gamma\gamma$ luminosity. Combining these two conditions, we obtained a 1D estimate of the laser parameters that optimize the $\gamma\gamma$ luminosity, $\hbar\omega_0=10$~eV and $c\tau=5$~mm, which we then tested against simulations with the code {\sc cain}.        

These simulations indicate that the choice of scattering laser wavelength is important for the future 10 TeV $\gamma\gamma$-collider concept. Near-optical lasers with photon energies of 2.5~eV and 5~eV and pulse lengths of approximately 6~mm maximize the partial $\gamma\gamma$ collision luminosities (see Fig.~\ref{fig:lumi scan}), whereas X-ray lasers, at 1~keV and 80~$\mu$m, generate sharply peaked luminosity spectra at high energies but at reduced total luminosity. The corresponding laser pulse energies are 125~J for the 5~eV case and 1.7~J for the 1~keV case. While for a 125 GeV Higgs factory a sharply peaked $\gamma\gamma$ luminosity spectrum is preferable, it has recently been shown \cite{chigusa.arxiv.2025,cipressi.arxiv.2026,fraser.arxiv.2026} that broad luminosity spectra do enable discovery capabilities for various collider concepts, including $\gamma\gamma$-colliders at multi-TeV energies. We note that although the optimized high-energy $\gamma\gamma$ luminosities ($L_{\gamma\gamma}^{5\%}$) are approximately 40 times smaller than the geometric luminosity of the colliding electron beams, the appropriate comparison is with the high-energy $e^-e^-$ and $e^+e^-$ luminosities obtained from beam-beam collision simulations, which account for beamstrahlung and pair production, among other effects. On that basis the difference is reduced to one order of magnitude (see Table~\ref{tab:collider lumi}).

We have shown that a future 10 TeV $\gamma\gamma$-collider can be designed with scattering lasers whose wavelength ranges from near-optical to X-ray with the former delivering higher luminosity and being less challenging from the technological point of view. The particular choice of the scattering laser wavelength will be guided by the scientific case for such a collider. The resulting high-energy photon luminosities indicate that a standalone $\gamma\gamma$-collider can enable a comprehensive physics program for the discovery of phenomena beyond the Standard Model. 

\section*{Acknowledgments}
This research was supported by the Laboratory Directed Research and Development Program of Lawrence Berkeley National Laboratory under U.S. Department of Energy Contract No.~DE-AC02-05CH11231. We thank Christiane Scherb, Inbar Savoray, Toby Opferkuch, Linda Xu, So Chigusa, and Simon Knapen for useful discussions and for collaboration on related work.

\section*{Appendix I: Polarization dependence}

\begin{figure}[!ht]
\centering
\includegraphics[width=8cm]{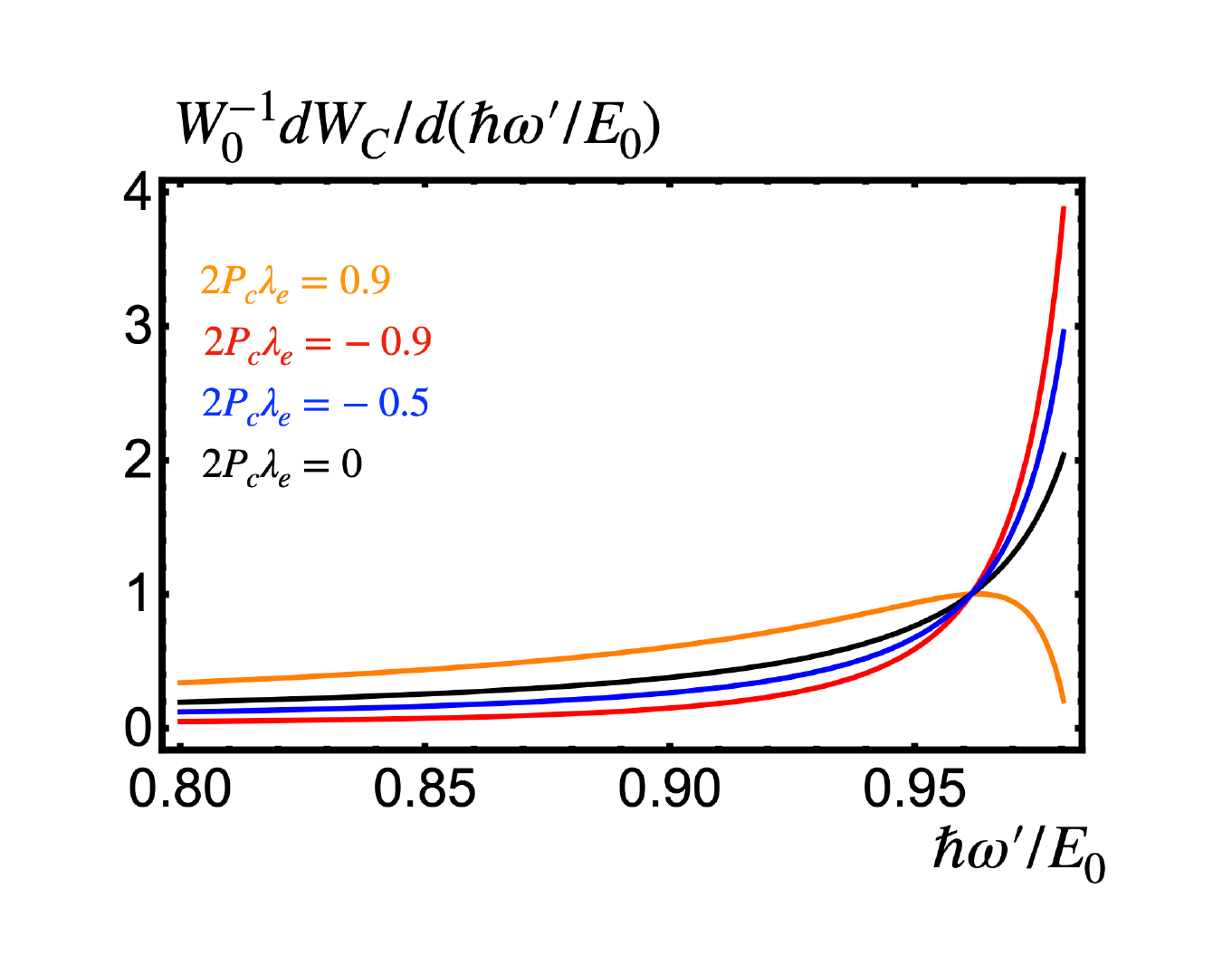}
\caption{Spectrum of photons after Compton scattering for different values of $2P_c\lambda_e$, at $x=50$.}
\label{fig:compton spectrum polarization}
\end{figure}

As noted above, the polarizations of the laser and electron beams play an important part in the design of a $\gamma\gamma$-collider. The Compton scattering rate can be written as \cite{tesla.ijmpa.2004}
\begin{equation}
    W_C=W_C^0+2\lambda_e P_c W_C^1,
\end{equation}
where $\lambda_e$ is the mean electron helicity ($|\lambda_e|\leq 1/2$) and $P_c$ is the mean helicity of the laser photons ($|P_c|\leq 1$), and where
\begin{equation}
    W_C^0=\frac{W_0}{x}\left[\left(1-\frac{4}{x}-\frac{8}{x^2}\right)\ln(x+1)+\frac{1}{2}+\frac{8}{x}-\frac{1}{2(x+1)^2}\right], \nonumber
\end{equation}
\begin{equation}
    W_C^1=\frac{W_0}{x}\left[\left(1+\frac{2}{x}\right)\ln(x+1)-\frac{5}{2}+\frac{1}{x+1}-\frac{1}{2(x+1)^2}\right], 
\end{equation}
with $W_0=\alpha a_0^2 \omega_0$ and $\alpha=1/137$ the fine structure constant. The photon spectrum (see Fig.~\ref{fig:compton spectrum polarization}) depends strongly on the polarizations of the initial laser and electron beams, especially in the high-energy region that is crucial for $\gamma\gamma$-collider performance.

\begin{figure}[!ht]
\centering
\includegraphics[width=8cm]{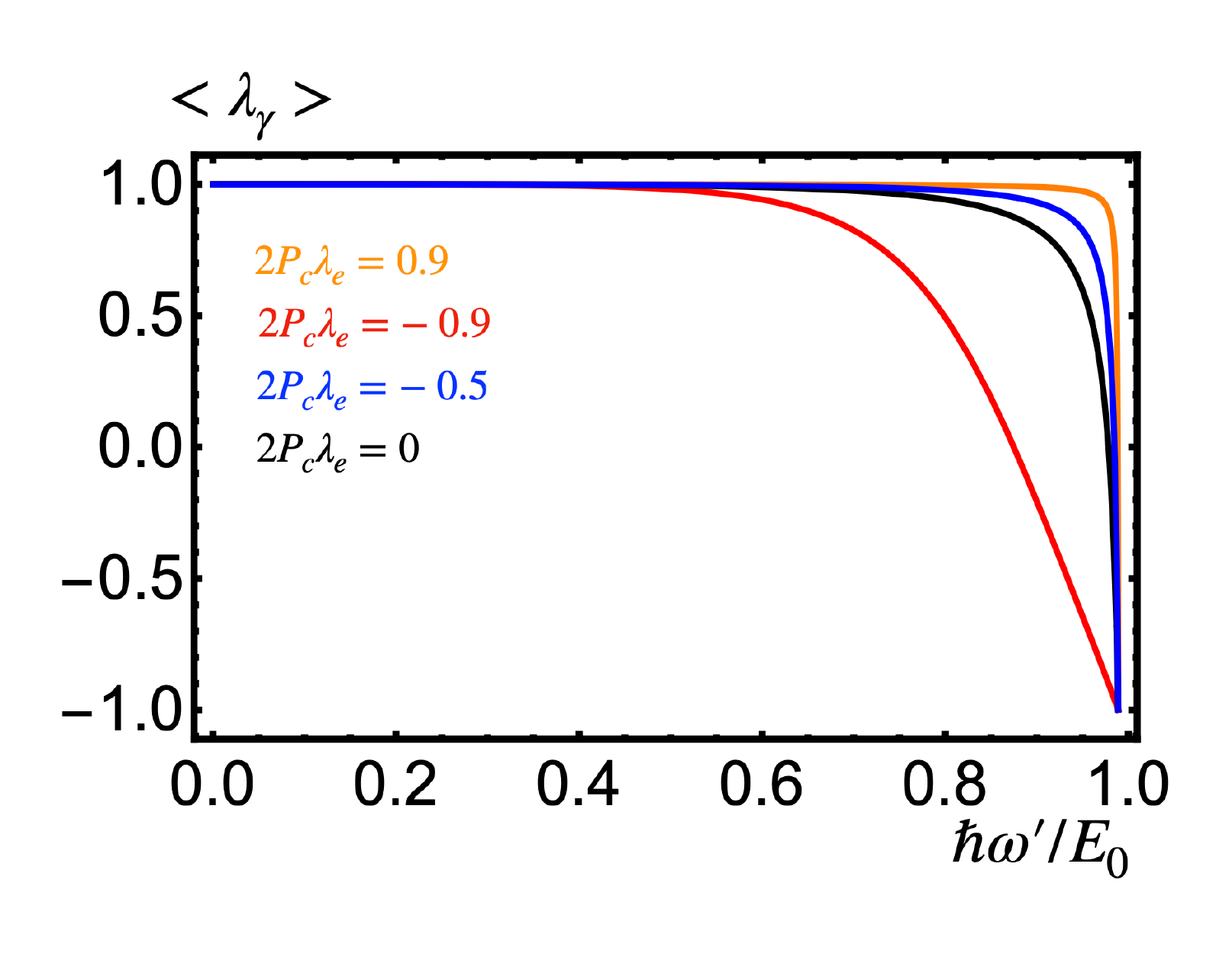}
\caption{Average helicity of photons emitted in Compton scattering for different values of $2P_c\lambda_e$, at $x=50$.}
\label{fig:helicity}
\end{figure}

The average helicity of a photon scattered by an electron is \cite{tesla.ijmpa.2004}
\begin{equation}
    \langle\lambda_\gamma\rangle=\frac{-P_c(2r-1)Y+2\lambda_e x r [1+(1-y)(2r-1)^2]}{Y-4r(1-r)-2\lambda_e P_c x r (2-y)(2r-1)},
\end{equation}
where $Y=(1-y)^{-1}+1-y$, $r=y(1-y)^{-1}x^{-1}$, and $y=\hbar\omega^\prime/E_0$. The dependence of the average photon helicity on the photon energy is shown in Fig.~\ref{fig:helicity} for different values of $2\lambda_e P_c$, in all cases with $P_c=1$. Low-energy photons are emitted with the same helicity as the laser pulse, whereas high-energy photons have the opposite helicity, and the extent of the region of opposite helicity depends on the value of $2\lambda_e P_c$, being largest when the laser and electron beams are initially oppositely polarized. Thus, as $2\lambda_e P_c$ approaches $-1$, the high-energy part of the photon spectrum develops a sharper peak and these photons are predominantly polarized in the same way.

\begin{figure}[!ht]
\centering
\includegraphics[width=8cm]{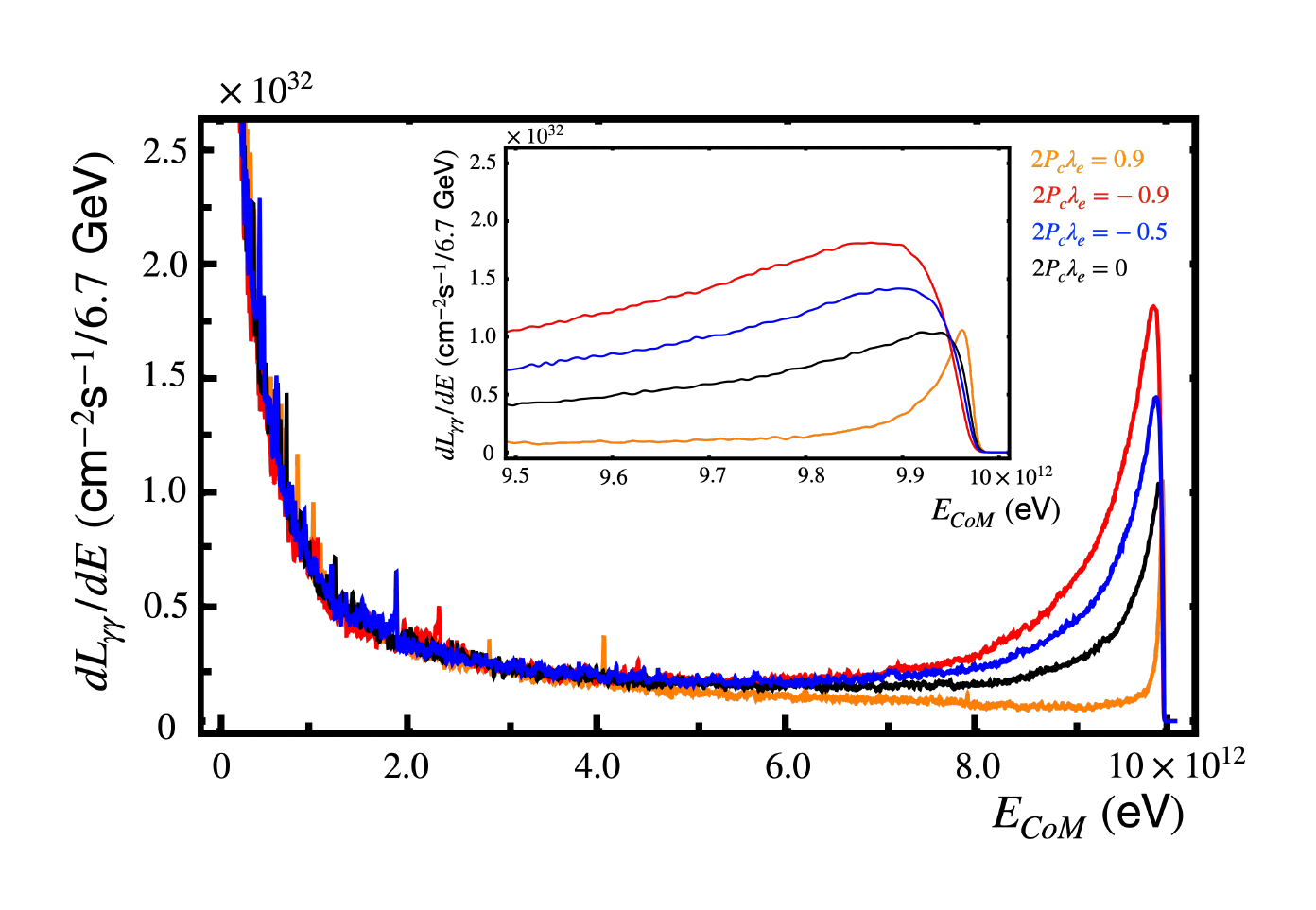}
\caption{The $\gamma\gamma$ luminosity spectra for $2P_c\lambda_e=0.9$ (orange), $-0.9$ (red), $-0.5$ (blue), and $0$ (black). The laser photon energy is $\hbar\omega_0=5$~eV.}
\label{fig:lumi pol}
\end{figure}

Since we are considering a $\gamma\gamma$-collider concept in which photon beam generation is accompanied by prolific $e^+e^-$ pair production, we now revisit the polarization dependence of the Breit-Wheeler process. Its rate can be written as \cite{tesla.ijmpa.2004}
\begin{equation}
    W_{BW}=W_{BW}^0+\lambda_\gamma P_c W_{BW}^1,
\end{equation}
where
\begin{eqnarray}
    W_{BW}^0=\frac{2 W_0}{x_\gamma}\left[2\left(1+\frac{4}{x_\gamma}-\frac{8}{x_\gamma^2}\right)\mbox{ArcTanh}\sqrt{1-\frac{4}{x_\gamma}} \right.\nonumber\\
    \left.-\left(1+\frac{4}{x_\gamma}\right)\sqrt{1-\frac{4}{x_\gamma}}\right],
\end{eqnarray}
\begin{equation}
    W_{BW}^1=-\frac{2W_0}{x_\gamma}\left[2\mbox{ArcTanh}\sqrt{1-\frac{4}{x_\gamma}} -3\sqrt{1-\frac{4}{x_\gamma}}\right],
\end{equation}
and $x_\gamma$ is defined analogously to $x$ as $x_\gamma=4\hbar\omega_0\hbar\omega^\prime/(m_e^2c^4)$. The cases $\lambda_\gamma P_c=+1$ and $\lambda_\gamma P_c=-1$ differ increasingly as $x_\gamma$ grows. For $x_\gamma<15$ the $\lambda_\gamma P_c=-1$ case is suppressed, meaning that the high-energy photons from the Compton process, which have $\lambda_\gamma P_c=-1$, are less likely to convert into $e^+e^-$ pairs than the low-energy photons, which have $\lambda_\gamma P_c=+1$. The situation reverses for $x_\gamma>15$, where the high-energy photons are the more likely to convert. The ratio of the $\lambda_\gamma P_c=-1$ and $\lambda_\gamma P_c=+1$ rates changes from $\sim 0.7$ at $x_\gamma=4.8$ to $\sim 2.3$ at $x_\gamma=10^5$. The polarization dependence of the Compton and Breit-Wheeler processes as discussed here neglects multiple emissions and realistic temporal and spatial laser profiles, and it cannot be reformulated directly to yield a polarization-sensitive expression for the $\gamma\gamma$ luminosity at the interaction point. To illustrate the dependence of the $\gamma\gamma$ luminosity spectrum on the polarizations of the initial electron and laser beams, we therefore performed {\sc cain} simulations of the $\gamma\gamma$-collider setup for different values of $2\lambda_e P_c$, taking the 5~eV laser case of Fig.~\ref{fig:lumi compare}. As noted above, all the simulations in the main part of the paper were carried out with $2\lambda_e P_c=-0.9$, corresponding to a realistic polarization fraction of 90\% for the accelerated electron beams; here we used $2\lambda_e P_c=0.9$, $-0.9$, $-0.5$, and $0$ (see Fig.~\ref{fig:lumi pol}). As expected, the case $2\lambda_e P_c=-0.9$ produces the highest peak of all the simulated cases in the high-energy part of the $\gamma\gamma$ luminosity spectrum, emphasizing the importance of choosing the polarizations of the electron and laser beams so as to maximize performance (see Table~\ref{tab:pol compare}). While the total $\gamma\gamma$ luminosity is almost the same for all polarization cases considered, the partial 5\% luminosity is six times larger for $2\lambda_e P_c=-0.9$ than for $2\lambda_e P_c=0.9$.        

\begin{table}
\caption{\label{tab:pol compare} Dependence of the $\gamma\gamma$ luminosity on the laser and electron beam polarizations, from Fig.~\ref{fig:lumi pol}.}
\begin{tabular}{c||c|c|}
\hline
$2\lambda_e P_c$ & $L_{\gamma\gamma}$, cm$^{-2}$ $s^{-1}$ & $L_{\gamma\gamma}^{5\%}$, cm$^{-2}$ $s^{-1}$ \\ \hline\hline
0.9 & $8.2\times 10^{34}$ & $1.4\times 10^{33}$  \\
-0.9 & $8.9\times 10^{34}$ & $8.4\times 10^{33}$  \\
-0.5 & $8.8\times 10^{34}$ & $6.3\times 10^{33}$  \\
0 & $8.2\times 10^{34}$ & $4.1\times 10^{33}$  \\
\hline
\end{tabular}
\end{table}

\section*{Appendix II: Dependence of the $\gamma\gamma$ luminosity on the electron bunch length}

The electron bunch length plays an important role in lepton collider design \cite{clic.arxiv.2018,ilc.arxiv.2022}. In what follows we explore how the $\gamma\gamma$-collider concept changes when longer electron bunches are used, which shows that the results of the present paper can be extended to other lepton collider designs.

\begin{figure}[!ht]
\centering
\includegraphics[width=8cm]{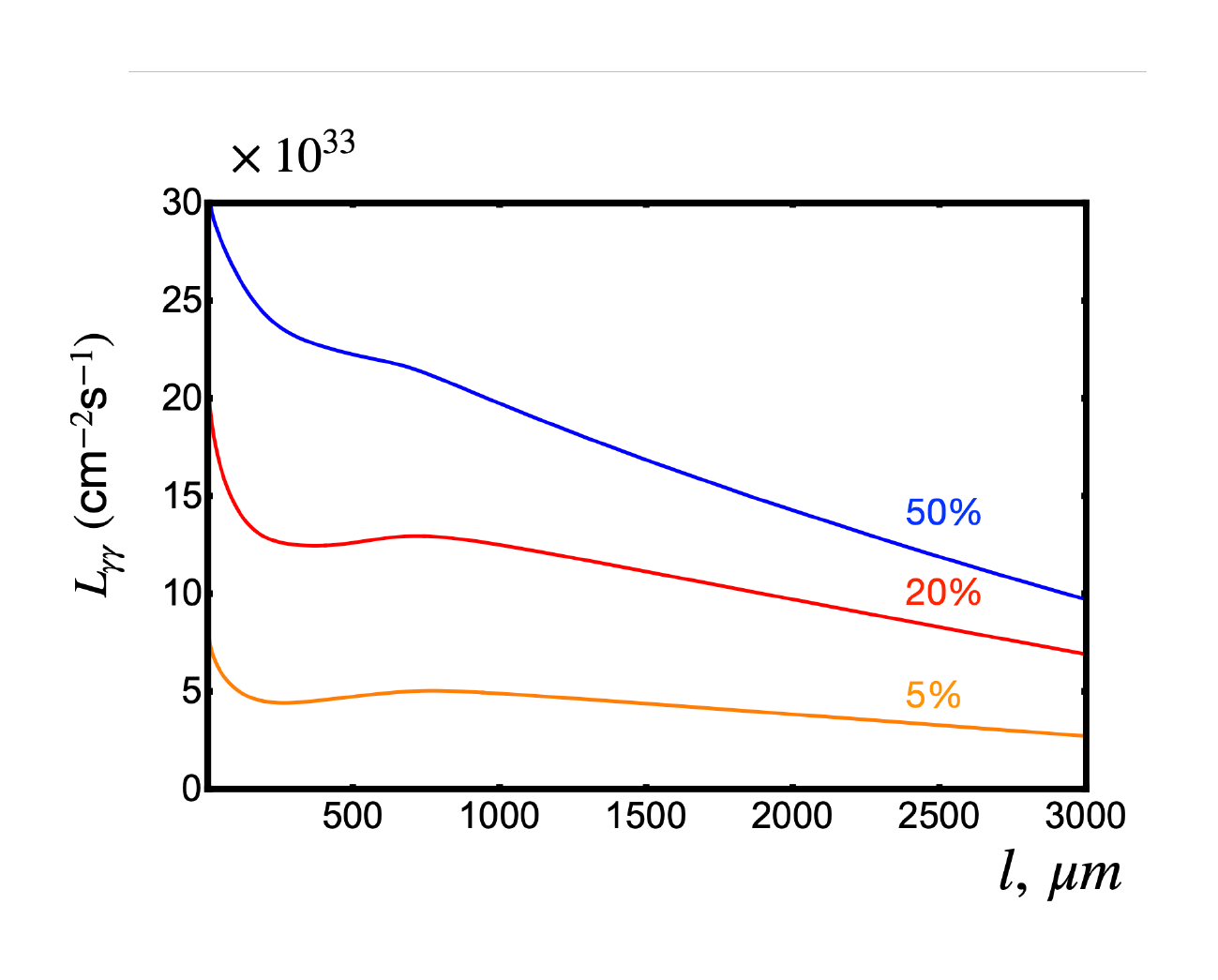}
\caption{Dependence of the partial $\gamma\gamma$ luminosities $L_{\gamma\gamma}^{5\%}$ (orange), $L_{\gamma\gamma}^{20\%}$ (red), and $L_{\gamma\gamma}^{50\%}$ (blue) on the electron bunch length, for a 5~eV scattering laser. The other interaction parameters are the same as in Fig.~\ref{fig:lumi compare}.}
\label{fig:length}
\end{figure}

Using the same parameters as in Fig.~\ref{fig:lumi compare}, we show in Fig.~\ref{fig:length} the dependence of the partial $\gamma\gamma$ luminosities on the electron bunch length $l$, varied from 5~$\mu$m (the value used in the {\sc cain} simulations above) to 3~mm. The number of high-energy photons generated at the conversion stage and their energy distribution are almost the same for all bunch lengths considered, so the differences in the partial luminosities must arise from the interaction of these photons with the colliding electron bunches and with the $e^+e^-$ pairs generated at the conversion stage. The partial luminosities fall sharply as the bunch length increases to tens of micrometers, and continue to decrease thereafter, but at a much slower rate. The maximum and minimum values in the figure differ by approximately a factor of three, while the bunch length increases by almost three orders of magnitude, from 5~$\mu$m to 3~mm. Several effects are responsible for this behavior. First, high-energy photons can convert into $e^+e^-$ pairs at the interaction point in the field of the opposing electron bunch, via the multi-photon Breit-Wheeler process, at the rate \cite{gonoskov.rmp.2022}
 \begin{equation}\label{gamma probability limits}
    W^{BW}=\begin{cases}
    0.073(\lambda_C/c)\frac{\alpha\chi_\gamma}{\hbar\omega^\prime/m c^2}\exp\left(-\frac{8}{3\chi_\gamma}\right),~~~\chi_\gamma\ll 1, \\
    0.67(\lambda_C/c)\frac{\alpha\chi_\gamma^{2/3}}{\hbar\omega^\prime/m c^2},~~~\chi_\gamma \gg 1,
    \end{cases}
    \end{equation}
where $\lambda_C = \hbar/(m c)$ is the reduced Compton wavelength of the electron. The parameter $\chi_\gamma$, which characterizes the interaction of a high-energy photon with the electromagnetic field of a 5~TeV electron bunch, can be written as
\begin{equation}
    \chi_\gamma=\frac{5}{6}\frac{r_e^2}{\alpha l}\frac{\hbar\omega^\prime}{m c^2}\frac{n_0}{\sigma_r},
\end{equation}
where $r_e$ is the classical electron radius and $\sigma_r$ is the electron bunch radius at the interaction point, or, equivalently, in terms of the geometric luminosity, as
\begin{equation}
    \chi_\gamma=\frac{5}{6}\frac{r_e^2}{\alpha l}\frac{\hbar\omega^\prime}{m c^2}\left(\frac{4\pi L_{geo}^{ee}}{f}\right)^{1/2}.
\end{equation}
A typical value for the parameters considered above is $\chi_\gamma\sim 10^3$. In a realistic collider this parameter will be reduced by electron beam energy loss at the conversion point, by beamstrahlung, and by disruption.  

The sharp drop in the partial luminosities can be attributed to the multi-photon Breit-Wheeler process, which is strongest when the electron bunch length approaches the radiation length of a photon in the field of the opposing bunch. When the bunch is longer than this radiation length, the main contribution to the luminosity comes from the front part of the photon pulse, over a length approximately equal to the radiation length: photons propagating further towards the back of each electron bunch have an increased probability of converting into $e^+e^-$ pairs and a significantly reduced chance of colliding with another high-energy photon. Adopting a simplified model of the photon interaction with the colliding electron bunches, one finds that the number of photons in this part scales as $l^{-1/3}$, and the luminosity as the square of that. In deriving this scaling we assumed that the photons are distributed homogeneously along the electron bunch, that they all experience the same field during the interaction (i.e., we neglected all beamstrahlung effects at the interaction point), and that the interaction takes place in the high-$\chi$ regime, where the multi-photon Breit-Wheeler rate is proportional to $\chi^{2/3}$.

\begin{figure}[!ht]
\centering
\includegraphics[width=8cm]{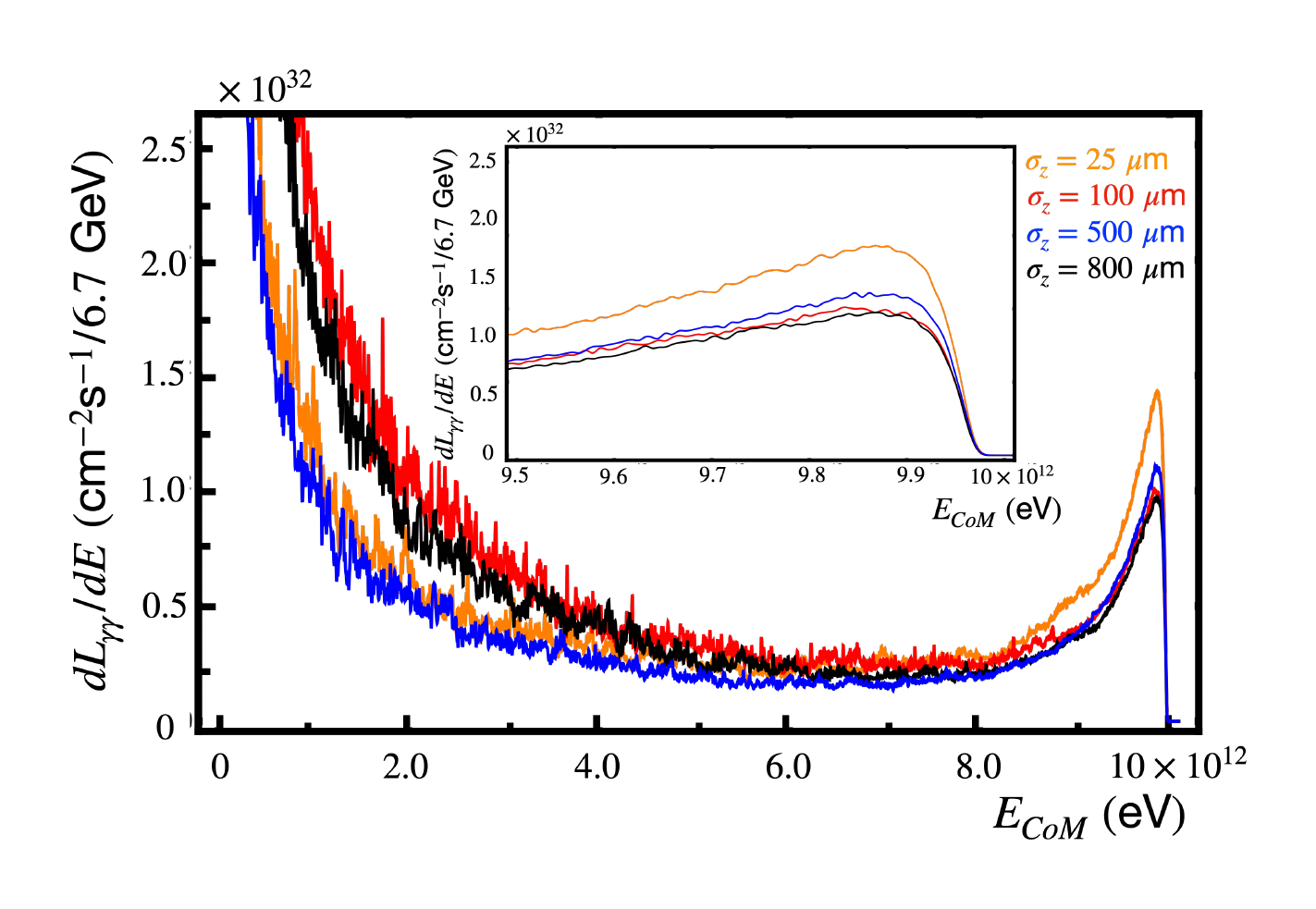}
\caption{The $\gamma\gamma$ luminosity spectra for different electron bunch lengths: $l=25$~$\mu$m (orange), $l=100$~$\mu$m (red), $l=500$~$\mu$m (black), and $l=800$~$\mu$m (blue). The inset shows the luminosity spectra near the 10~TeV peak. The other interaction parameters are the same as in Fig.~\ref{fig:lumi compare}.}
\label{fig:spectra_length}
\end{figure}

The {\sc cain} simulations (Fig.~\ref{fig:length}) show a more complicated picture. Although the predicted trends are present, one must also take into account the disruption of the electron bunches at the interaction point, which becomes stronger with increasing bunch length; the different scaling of the multi-photon Breit-Wheeler rate with $\chi$, since the interaction is no longer in the high-$\chi$ regime; and the change in the photon bunch radius at the interaction point once the bunch length becomes comparable to the scattering laser length. We note that the $\gamma\gamma$ luminosity spectra for different electron bunch lengths are very similar, as shown in Fig.~\ref{fig:spectra_length}, with the exception of the shortest length, which has a higher peak at high photon energies. The differences in the partial and total $\gamma\gamma$ luminosities are small: for example, $L_{\gamma\gamma}(l=25~\mu\mbox{m})=1.2\times 10^{35}$~cm$^{-2}$s$^{-1}$ and $L_{\gamma\gamma}^{5\%}(l=25~\mu\mbox{m})=6.6\times 10^{33}$~cm$^{-2}$s$^{-1}$, whereas $L_{\gamma\gamma}(l=500~\mu\mbox{m})=1.6\times 10^{35}$~cm$^{-2}$s$^{-1}$ and $L_{\gamma\gamma}^{5\%}(l=500~\mu\mbox{m})=4.6\times 10^{33}$~cm$^{-2}$s$^{-1}$.

We conclude that these preliminary results indicate that the use of longer electron bunches does not invalidate the $\gamma\gamma$-collider concept, which makes it possible to extend the findings of this paper to $\gamma\gamma$-colliders based on other accelerator types.

\section*{Appendix III: High geometric luminosity}


It is worth considering a $\gamma\gamma$-collider design with a higher geometric luminosity, in order to determine whether this translates into a more efficient $\gamma\gamma$ collision rate at the IP than in the lower geometric luminosity cases. The most straightforward solution would be to increase the collider repetition rate, which should not affect this efficiency; using higher-charge electron bunches may be more effective. We do not consider here whether wakefield acceleration is able to generate such bunches. In what follows we use electron bunches with a charge of $N_e=6.3\times 10^9$ ($\sim 1$~nC) and an IP beta function of $\beta=0.34$~mm, which gives the minimum geometric spot size for $\gamma\epsilon=50$~nm and $E_0=5$~TeV once the Oide effect \cite{oide.prl.1988,blanco.prab.2016} is taken into account. With these two modifications the geometric luminosity is $L_{e^-e^-}^{geo}=6.0 \times 10^{36}$~cm$^{-2}$s$^{-1}$, at least an order of magnitude higher than the values of $5.5 \times 10^{35}$ and $3.3 \times 10^{35}$~cm$^{-2}$s$^{-1}$ considered above.

\begin{figure}[!ht]
\centering
\includegraphics[width=8cm]{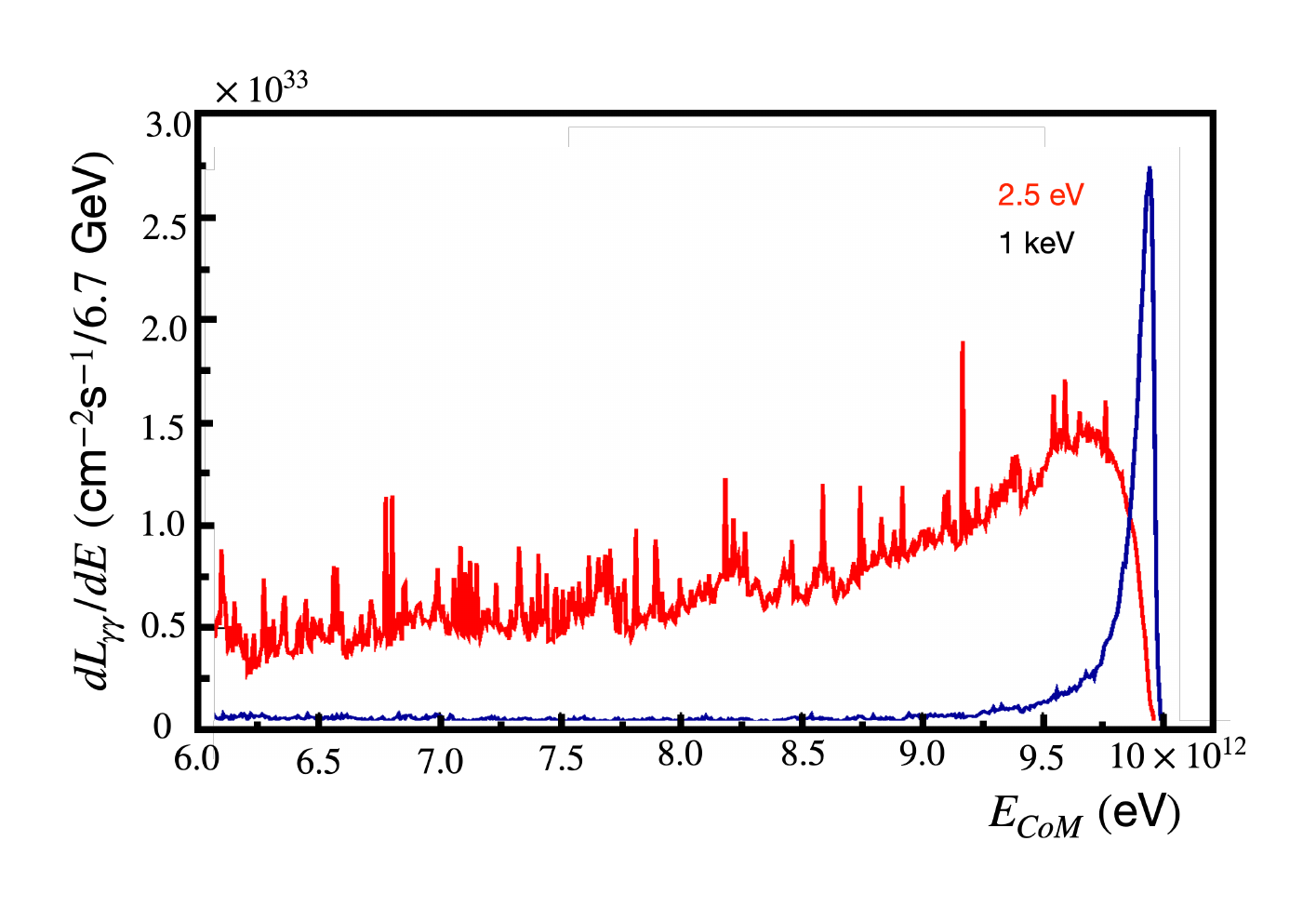}
\caption{The $\gamma\gamma$ luminosity spectra for $L_{e^-e^-}^{geo}=6.0 \times 10^{36}$~cm$^{-2}$s$^{-1}$, with $\hbar\omega_0=2.5$~eV (red) and 1~keV (blue).}
\label{fig:lumi_high_lgeo}
\end{figure}

We found that the choice of scattering laser wavelength is important here as well. While particular choices of the near-optical (5~eV) and X-ray (1~keV) laser durations maximize the partial $\gamma\gamma$ collision luminosities, the two types of laser produce quite different $\gamma\gamma$ luminosity spectra (see, for example, Fig.~\ref{fig:lumi compare}b for $L_{e^-e^-}^{geo}=3.3 \times 10^{35}$~cm$^{-2}$s$^{-1}$, Fig.~\ref{fig:lumi} for $5.5 \times 10^{35}$~cm$^{-2}$s$^{-1}$, and Fig.~\ref{fig:lumi_high_lgeo} for $6.0 \times 10^{36}$~cm$^{-2}$s$^{-1}$). The shape of the spectra obtained with either near-optical or X-ray lasers remains similar for the different values of the geometric luminosity. The ratio of the 5\% partial luminosity to the geometric luminosity, $L^{5\%}_{\gamma\gamma}/L_{e^-e^-}^{geo}$, is however reduced as $L_{e^-e^-}^{geo}$ increases, from 0.025 to 0.015 for the 5~eV laser and from 0.01 to 0.008 for the 1~keV laser. The width of the high-energy peak in the luminosity spectra depends only weakly on the geometric luminosity, remaining at $\Delta E_{CoM}/E_{CoM}\approx0.05$ for the 5~eV laser and decreasing from 0.008 to 0.0065 for the 1~keV laser. As the geometric luminosity is increased by raising the bunch charge and focusing more tightly at the IP, however, the amount of beamstrahlung and coherent $e^+e^-$ pair production also increases. This should produce a larger background of low-energy photons, electrons, and positrons, the effect of which on the science reach of a 10 TeV $\gamma\gamma$-collider should be addressed in future work.

\section*{Appendix IV: Electron beam and laser parameters used in the {\sc cain} simulations}

The simulations of the $\gamma\gamma$-collider concept were carried out with the specialized Monte Carlo code {\sc cain} \cite{slaclab_cain,yokoya_cain_manual}, which was designed to study beam-beam interactions and laser-particle physics in high-energy linear colliders. The code uses a macro-particle approach to describe the behavior of electron, positron, and photon bunches, including their polarization, in the presence of self-generated and external electromagnetic fields and of QED processes. These processes include nonlinear and multi-photon Compton scattering ($e+n\gamma_L \rightarrow e + \gamma$), i.e., the interaction of high-energy electrons with an intense laser field or with the collective field of an opposing electron or positron bunch; bremsstrahlung ($e+e\rightarrow e+e+\gamma$); and coherent and incoherent $e^+e^-$ pair production through the nonlinear and multi-photon Breit-Wheeler ($\gamma+n\gamma_L \rightarrow e^+ +e^-$), Bethe-Heitler ($\gamma + e^\pm \rightarrow e^\pm + e^+ + e^-$), and Landau-Lifshitz ($e + e \rightarrow e + e + e^+ + e^-$) processes. The Bethe-Heitler, Landau-Lifshitz, and bremsstrahlung processes are calculated using the equivalent photon approximation \cite{berestetsky.1982}, whereas the multi-photon Compton and Breit-Wheeler processes are treated within the local constant field approximation (see Ref.~\cite{gonoskov.rmp.2022} and references therein for details of this approximation).

The simulation results reported in this paper were obtained with a two-stage {\sc cain} setup. First, the collision of the electron beams with the focused laser pulses was modeled, yielding two counter-propagating beams of electrons, positrons, and photons, with each particle characterized by its position, momentum, and polarization. These beams were then used to initialize the beam-beam collision simulation at the IP, which constitutes the second stage. The output consists of 1D and 2D luminosity spectra for the various collision channels: $\gamma\gamma$, $e^-\gamma$, $e^-e^-$, $e^+\gamma$, and $e^+e^-$.

The parameters of the $\gamma\gamma$, $e^-e^-$, and $e^+e^-$ collider simulations discussed in the main part of the paper are summarized in Tables~\ref{tab:beam parameters low} and \ref{tab:beam parameters high}. Figures~\ref{fig:lumi compare}, \ref{fig:lumi a0}, and \ref{fig:lumi pol}-\ref{fig:spectra_length} use the same parameters as Table~\ref{tab:beam parameters low}, except that the beta function was set to $\beta=1.0$~mm, giving $L_{ee}^{geo}=3.3\times 10^{35}$~cm$^{-2}$s$^{-1}$. The parameters used for Fig.~\ref{fig:lumi_high_lgeo} in Appendix III are listed in Table~\ref{tab:beam parameters high}; they differ from those of Table~\ref{tab:beam parameters low} in the use of a higher-charge, more tightly focused electron beam. 

\begin{table}
\caption{\label{tab:beam parameters low} Electron beam and laser pulse parameters used to generate Figs.~\ref{fig:lumi} and \ref{fig:lumi scan}.}
\begin{tabular}{|cc||cc|}
\hline
\multicolumn{2}{|c||}{electron beam}    & \multicolumn{2}{c|}{laser pulse}    \\ \hline\hline
\multicolumn{1}{|l|}{$N_e$}  & $2.12\times 10^9$ & \multicolumn{1}{l|}{$a_0$} & $0.3$ \\ \hline
\multicolumn{1}{|l|}{$\sigma_z$} & $5~\mu$m & \multicolumn{1}{l|}{$\lambda_0$} & 2 $\mu$m -- 0.25 nm \\ \hline
\multicolumn{1}{|l|}{$f$} & $47$ kHz & \multicolumn{1}{l|}{$\hbar\omega_0$} & 0.65 eV -- 5 keV \\ \hline
\multicolumn{1}{|l|}{$E_0$} & 5 TeV & \multicolumn{1}{l|}{$P_C$} & 1 \\ \hline
\multicolumn{1}{|l|}{$\epsilon$} & $50/\gamma$ nm & \multicolumn{1}{l|}{$\tau$} & 60 fs -- 40 ps \\ \hline
\multicolumn{1}{|l|}{$\beta$} & 0.6 mm & \multicolumn{1}{l|}{$c\tau$} & 20 $\mu$m -- 12 mm \\ \hline
\multicolumn{1}{|l|}{$L_{ee}^{geo}$} & $5.5\times 10^{35}$ cm$^{-2}$s$^{-1}$ & \multicolumn{1}{l|}{$w_0$} & $20\lambda_0$ \\ \hline
\end{tabular}
\end{table}

\begin{table}
\caption{\label{tab:beam parameters high} Electron beam and laser pulse parameters used to generate Fig.~\ref{fig:lumi_high_lgeo}.}
\begin{tabular}{|cc||cc|}
\hline
\multicolumn{2}{|c||}{electron beam}    & \multicolumn{2}{c|}{laser pulse}    \\ \hline\hline
\multicolumn{1}{|l|}{$N_e$}  & $6.3\times 10^9$ & \multicolumn{1}{l|}{$a_0$} & $0.3$ \\ \hline
\multicolumn{1}{|l|}{$\sigma_z$} & $5~\mu$m & \multicolumn{1}{l|}{$\lambda_0$} & 0.5 $\mu$m and 1.24 nm \\ \hline
\multicolumn{1}{|l|}{$f$} & $47$ kHz & \multicolumn{1}{l|}{$\hbar\omega_0$} & 2.5 eV and 1 keV \\ \hline
\multicolumn{1}{|l|}{$E_0$} & 5 TeV & \multicolumn{1}{l|}{$P_C$} & 1 \\ \hline
\multicolumn{1}{|l|}{$\epsilon$} & $50/\gamma$ nm & \multicolumn{1}{l|}{$\tau$} & 20 ps and 267 fs \\ \hline
\multicolumn{1}{|l|}{$\beta$} & 0.34 mm & \multicolumn{1}{l|}{$c\tau$} & 6.0 mm and 80 $\mu$m \\ \hline
\multicolumn{1}{|l|}{$L_{ee}^{geo}$} & $6.0\times 10^{36}$ cm$^{-2}$s$^{-1}$ & \multicolumn{1}{l|}{$w_0$} & $20\lambda_0$ \\ \hline
\end{tabular}
\end{table}

\newpage

\bibliography{refs.bib}

\end{document}